\documentclass[reprint,showpacs,showkeys,superscriptaddress,longbibliography,floatfix,10pt]{revtex4-2}
\usepackage{indentfirst}
\usepackage{bm}
\usepackage{dcolumn}
\usepackage{amssymb}
\usepackage{amsmath}
\usepackage{graphicx}
\usepackage{dcolumn}
\usepackage{bm}
\usepackage{xcolor}
\usepackage{fix-cm}
\usepackage{mathptmx}
\usepackage[T1]{fontenc}
\usepackage{epstopdf}
\usepackage{longtable}
\usepackage[colorlinks,citecolor=blue,linkcolor=blue,anchorcolor=blue,filecolor=blue,urlcolor=blue]{hyperref}
\usepackage{multirow}
\usepackage{slashed}
\makeatletter
\def\NAT@def@citea{\def\@citea{\NAT@separator}}
\makeatother

\begin{document}	
	\hyphenpenalty=5000
	\tolerance=2000
	
		\title{Tetrahedral shape and Lambda impurity effect in $^{80}$Zr with a multidimensionally constrained relativistic Hartree-Bogoliubov model}
		
		\author{Dan Yang}
		\affiliation{%
			Department of Physics, Guangxi Normal University,%
			Guilin,%
			541004,%
			China}
		\affiliation{Guangxi Key Laboratory of Nuclear Physics and Technology, Guangxi Normal University, Guilin, 541004, China}

		\author{Yu-Ting Rong}
		\email[Corresponding author:~]{rongyuting@gxnu.edu.cn}
		\affiliation{%
	Department of Physics, Guangxi Normal University,%
	Guilin,%
	541004,%
	China}
		\affiliation{Guangxi Key Laboratory of Nuclear Physics and Technology, Guangxi Normal University, Guilin, 541004, China}
	
		
		\begin{abstract}

               This study investigates the tetrahedral structure in $^{80}$Zr and Lambda ($\Lambda$) impurity effect in $^{81}_{~\Lambda}$Zr using the multidimensionally constrained relativistic Hartree-Bogoliubov model.
       The ground states of both $^{80}$Zr and $^{81}_{~\Lambda}$Zr exhibit a tetrahedral configuration, accompanied by prolate and axial-octupole shape isomers. 
       Our calculations reveal there are changes in the deformation parameters $\beta_{20}$, $\beta_{30}$, and $\beta_{32}$ upon $\Lambda$ binding to $^{80}$Zr, except for $\beta_{32}$ when $\Lambda$ occupies $p$-orbits. 
       Compared to the two shape isomers, the $\Lambda$ particle exhibits weaker binding energy in the tetrahedral state when occupying the $1/2^+[000](\Lambda_s)$ or $1/2^-[110]$ single-particle states. In contrast, the strongest binding occurs for the $\Lambda$ particle in the $1/2^-[101]$ state with tetrahedral shape. Besides, a large $\Lambda$ separation energy may not necessarily correlate with a significant overlap between the density distributions of the $\Lambda$ particle and the nuclear core, particularly for tetrahedral hypernuclei.

      \end{abstract}
		

		\maketitle
	
\section{INTRODUCTION}\label{sec1}

Tetrahedral symmetry, a prevalent arrangement in molecules and metallic clusters, can also manifest in nuclei. This specific symmetry arises from nonaxial octupole deformations $\beta_{32}$. 
Similar to the well-established concept of magic numbers for spherical and prolate/oblate nuclei, specific configurations of protons and neutrons that fill full shells are predicted to be associated with enhanced stability for nuclei adopting a tetrahedral shape. 
These configurations effectively close certain energy shells, leading to tighter binding energy for the nucleus.
Consequently, nuclei with specific neutron (proton) numbers, $N(Z) = 16,~20,~32,~40,~56-58,~70,~90-94$ and $N = 112,~136/142$, are predicted to have deformed gap sizes comparable, or larger than the strongest spherical gaps at $Z=20$, 28, 40, or 50~\cite{Dudek2002_PRL88-252502}.

Among nuclei predicted to exhibit a tetrahedral shape, $^{80}$Zr serves as a prime example. Previous theoretical calculations predict a low-energy tetrahedral configuration for $^{80}$Zr alongside its known prolate ground state~\cite{Takami1998_PLB431-242,Yamagami2001_NPA693-579}.  Later, the Hartree-Fock-Bogoliubov
calculations with Gogny interations found the ground state
of $^{80}$Zr is tetrahedral carried by pure $\beta_{32}$ deformation~\cite{Tagami2014_PS89-054013,Tagami2015_JPG42-015106}. However, experimental observations based on decay properties indicate a highly deformed, non-tetrahedral shape for this nucleus~\cite{Lister1987_PRL59-1270,Llewellyn2020_PRL124-152501}. 
{Recent mass measurement  found it is more strongly bound than predicted and might be classified as a deformed doubly magic system~\cite{Hamaker2021_NP17-1408}.} However, this observation does not resolve the discrepancy regarding its shape.
Tagami et al~\cite{Tagami2014_PS89-054013} proposed that the observed "superdeformed" band in $^{80}$Zr might be an excited state, while the tetrahedral ground state remains elusive. 

Hypernucleus is composes of nucleons and hyperons.
Due to the additional strangeness degree of freedom, hyperons can penetrate deeply into the nucleus, acting as sensitive probes of various nuclear properties. These include effects like the shrinkage of size~\cite{Hiyama1999_PRC59-2351,Tanida2001_PRL86-1982}, the modification of cluster~\cite{Isaka2011_PRC83-044323,Isaka2015_PRC92-044326,Tanimura2019_PRC99-034324} and halo structures~\cite{Hiyama1996_PRC53-2075,Lu2002_CPL19-1775,Lu2003_EPJA17-19,Xue2022_PRC106-044306,Zhang2022_PTEP2022-023D01}, and the enhancement of the pseudospin symmetry in the nucleons~\cite{Lu2017_JPG44-125104}. Additionally, hypernuclei have been linked to the extension of the nuclear drip line~\cite{Lu2003_EPJA17-19,Samanta2008_JPG35-065101,Zhou2008_PRC78-054306,Wirth2018_PLB779-336}, the increase of fission barrier heights~\cite{Minato2009_NPA831-150,Minato2011_NPA856-55}, and the modification of shapes.
{These hyperon impurity effects are associate with nuclear shapes and the hyperon single-particle levels.}
In axial and reflection-symmetric nuclei, a $\Lambda$ hyperon occupying the lowest $s$-state (represented by $\Lambda_s$ or Nilsson quantum numbers $\Omega^\pi[N n_3 m_l]=1/2^+[000]$) reduces the quadrupole deformation parameter $\beta_2$, effectively shrinking the nucleus~\cite{Zhou2007_PRC76-034312,Win2008_PRC78-054311,Lu2014_PRC89-044307,Rong2021_PRC104-054321}. Conversely, a $\Lambda$ hyperon occupying the $1/2^-[110]$ Nilsson orbit derived from the $p$ shell ($\Lambda_p$) makes the nucleus more prolate, while those occupying the nearly degenerate $3/2^-[101]$ and $1/2^-[101]$ orbits make it more oblate\cite{Win2008_PRC78-054311,Isaka2011_PRC83-044323,Li2018_PRC97-034302,Chen2021_SciChinaPMA64-282011,Chen2022_CPC46-064109}. 
In triaxial calculations, a $\Lambda$ hyperon can soften the potential energy surface (PES) along the $\gamma$-direction (Hill-Wheeler coordinate)~\cite{Lu2011_PRC84-014328,Win2011_PRC83-014301,Yao2011_NPA868-12}. This alters the $2^+$ excitation energy and $E_2$ transition probability of the core nucleus~\cite{Yao2011_NPA868-12,Isaka2013_PRC87-021304R,Xue2015_PRC91-024327,Cui2015_PRC91-054306}. {Additionally, $\Lambda$ hyperon can split the rotational bands of the corresponding nucleus when it occupys $p$-orbits~\cite{Isaka2013_PRC87-021304R}.}
In calculations assuming axial symmetry but reflection asymmetry, with increasing octupole deformation, the additional $\Lambda_s (\Lambda_p)$ hyperon becomes more concentrated around the bottom (top) of the pear-shaped nucleus~\cite{Xia2019_SciChinaPMA62-042011}. However, there is no further study concerning the exotic tetrahedral hypernuclei up to now.

Therefore, this work investigates the shape of $^{80}$Zr and the $\Lambda$ hyperon impurity with tetrahedral shapes. Although it is difficult to synthesis the hypernucleus $^{81}_{\Lambda}$Zr, this work will give a theoretical understanding for hyperon impurity effect with $Y_{32}$ correlation. We employ the multidimensionally-constrained covariant density functional theories (MDC-CDFTs), where the shape degrees of freedom (represented by $\beta_{\lambda\mu}$ with even $\mu$ values) are self-consistently included under the intrinsic $V_4$ symmetry group~\cite{Lu2012_PRC85-011301R,Lu2014_PRC89-014323,Zhou2016_PS91-063008,Zhao2017_PRC95-014320}. This framework has been successfully applied to study tetrahedral shapes in neutron-rich Zr isotopes~\cite{Zhao2017_PRC95-014320}, $Y_{32}$ correlations in $N=150$ isotones~\cite{Zhao2012_PRC86-057304}, and extended to hypernuclear properties~\cite{Lu2011_PRC84-014328,Lu2014_PRC89-044307,Rong2020_PLB807-135533,Rong2021_PRC104-054321,Chen2021_SciChinaPMA64-282011}.
In the following, we will introduce the MDC-CDFTs for hypernuclei in Sec.~\ref{sec:model}. In Sec.~\ref{sec:results}, we will use MDC-CDFTs to calculate the potential energy surfaces (PESs) of $^{80}$Zr and $^{81}_\Lambda$Zr and discuss their shapes. The impurity effect in a tetrahedral nucleus is analyzed based on $\Lambda$ separation energies, charge radii, and density distributions. Finally, a summary is given in Sec.~\ref{sec:summary}.

\section{THEORETICAL FRAMEWORK}\label{sec:model}

The CDFTs are microscopic nuclear models, which have been very successful in describing properties of nuclear matter
and finite nuclei~\cite{Reinhard1989_RPP52-439,Ring1996_PPNP37-193,Bender2003_RMP75-121,Vretenar2005_PR409-101,Meng2006_PPNP57-470,Niksic2011_PPNP66-519,Liang2015_PR570-1,Meng2015_JPG42-093101,Zhou2016_PS91-063008,Meng2016_RDFNS}.
In CDFTs, baryons are interact through mesons, and the Lagrangian for a $\Lambda$ hypernucleus is written as

\begin{widetext}
\begin{equation}\label{eq:Lagrangian}
	\begin{split}
		{\cal L} = & \sum_{B}\bar{\psi}_B
		\bigg(
		i \gamma_\mu \partial^\mu - M_B 
		- g_{\sigma B} \sigma  
		- g_{\omega B} \gamma_\mu \omega^\mu    
		- g_{\rho B}   \gamma_\mu \vec{\tau} \cdot \vec{\rho}^\mu        
		- e \gamma_\mu \dfrac{1-\tau_3}{2} A^\mu
		\bigg) \psi_B 
		+ \psi_\Lambda\dfrac{f_{\omega\Lambda\Lambda}}{4M_\Lambda} \sigma_{\mu\nu} \Omega^{\mu\nu}\psi_\Lambda \\
		&+ \dfrac{1}{2} \partial^\mu \sigma   \partial_\mu \sigma   - \dfrac{1}{2} m_\sigma^2   \sigma^2               
		- \dfrac{1}{4} \Omega^{\mu\nu} \Omega_{\mu\nu} + \dfrac{1}{2} m_\omega^2 \omega^\mu \omega_\mu 
		- \dfrac{1}{4} \vec{R}^{\mu\nu} \vec{R}_{\mu\nu} 
		+ \dfrac{1}{2} m_\rho^2 \vec{\rho}^\mu \vec{\rho}_\mu 
		- \dfrac{1}{4} F^{\mu\nu}F_{\mu\nu}         
		,
	\end{split}
\end{equation}
\end{widetext}
where $B$ represents baryon (neutron, proton or $\Lambda$), and $M_B$ is the corresponding mass. 
$\sigma$, $\omega^\mu$ and $\vec{\rho}^\mu$ are scalar-isoscalar, vector-isoscalar and vector-isovector meson fields coupled to baryons, respectively.
$A^\mu$ is the photon field.  
$\Omega_{\mu\nu}$, $\vec{R}_{\mu\nu}$, and $F_{\mu\nu}$ are field 
tensors of the vector mesons $\omega^\mu$, $\vec{\rho}^\mu$, 
and photons $A^\mu$. 
$m_\sigma$ ($g_{\sigma B}$), 
$m_\omega$ ($g_{\omega B}$) and $m_\rho$ ($g_{\rho B}$) 
are the masses (coupling constants) for meson fields. 
Note that $\sigma^*$ and $\phi$ mesons introduced for multihypernuclei~\cite{Schaffner1994_AP235-35,Shen2006_PTP115-325,Rong2020_PLB807-135533} and nuclear matter~\cite{Schaffner1996_PRC53-1416,Weissenborn2013_NPA914-421,Tu2022_ApJ925-16} are omitted in this work because we only focus on single-$\Lambda$ hypernuclei.

The nonlinear coupling terms for mesons~\cite{Boguta1977_NPA292-413, Sugahara1994_NPA579-557,Long2004_PRC69-034319} and density dependence of the coupling constants~\cite{Typel1999_NPA656-331,Niksic2002_PRC66-024306,Lalazissis2005_PRC71-024312} were two different ways that introduced to give proper saturation properties of the nuclear matter, and extended to the investigation of hypernuclei. 
For nonlinear coupling effective interactions, nonlinear meson fields are added to the Lagrangian in Eq.~(\ref{eq:Lagrangian})~\cite{Shen2006_PTP115-325,Wang2013_CTP60-479,Sun2018_CPC42-25101}. For density-dependent effective interaction, the coupling constants are dependent on the total baryonic density $\rho^\upsilon$ as 
\begin{equation}
	g_{mB}(\rho^\upsilon)=g_{mB}(\rho_{\rm sat})f_{mB}(x), \quad x=\rho^\upsilon/\rho_{\rm sat},
\end{equation}
where $m$ labels mesons, $\rho_{\rm sat}$ is the saturation density of nuclear matter, $f_{mB}$ describes the density dependence behaviour~\cite{van-Dalen2014_PLB734-383,Rong2021_PRC104-054321}.
Since nonlinear coupling methods are widely used, we only introduce the detail framework adopting density-dependent coupling constants next.

Starting from the Lagrangian (\ref{eq:Lagrangian}) with density-dependent coupling constants, the equations of motion 
can be derived via the variational principle.
The Dirac equation for baryons reads 
\begin{eqnarray}\label{Dirac_equ}
	&  h_B       \psi_{iB}
	=  
	\varepsilon_i \psi_{iB},
	\label{eq:Dirac}
\end{eqnarray}
where $\varepsilon_i$ the single-particle energy, $\psi_{iB}$ the single-particle wave function, and the single-particle hamiltonian 
\begin{equation}
	h_B= \bm{\alpha} \cdot \bm{p} + V_B + T_B + \Sigma_R + \beta(M_B + S_B).
\end{equation}
The Klein-Gordon equations for mesons and the Proca equation for photon are
\begin{equation}
	\begin{split}
		(-\Delta + m_\sigma^2)     \sigma    &= -g_{\sigma N}         \rho_N^s        - g_{\sigma\Lambda} \rho_\Lambda^s, \\
		(-\Delta + m_\omega^2)     \omega_0  &=  g_{\omega N}         \rho_N^\upsilon + g_{\omega\Lambda} \rho_\Lambda^\upsilon    
		-\dfrac{f_{\omega\Lambda\Lambda}}{2M_\Lambda} \rho_\Lambda^T, \\
		(-\Delta + m_\rho^2)       \rho_0    &=  g_{\rho N}          (\rho_n^\upsilon - \rho_p^\upsilon), \\
		-\Delta A_0                         &=  e                    \rho_p^\upsilon.
	\end{split}
	\label{eq:KG}
\end{equation}
The Eq.~(\ref{eq:Dirac}) and Eqs.~(\ref{eq:KG}) 
are coupled via the scalar, vector, and tensor densities
\begin{equation}\label{eq:densities}
	\begin{split} 
		\rho_B^s        & =  \sum_i \bar{\psi}_{iB}          \psi_{iB}, \\ 
		\rho_B^\upsilon & =  \sum_i \bar{\psi}_{iB} \gamma^0 \psi_{iB}, \\ 
		\rho_\Lambda^T  & = i \bm{\partial} 
		\left(
		\sum_i \psi_{i\Lambda}^\dag \bm{\gamma} \psi_{i\Lambda}
		\right),
	\end{split}
\end{equation}
and various potentials
\begin{equation}\label{eq:potentials}
	\begin{split}
		V_B        = &~  g_{\omega B}\omega_0               
		+ g_{\rho B}\tau_3 \rho_0      + e\dfrac{1-\tau_3}{2}A_0,\\
		S_B        = &~  g_{\sigma B}\sigma           , \\
		T_\Lambda  = &~- \dfrac{ f_{\omega\Lambda\Lambda} }{ 2M_\Lambda } \beta(\bm{\alpha}\cdot \bm{p}) \omega_0, \\
		\Sigma_R   = &~  \dfrac{ \partial g_{\sigma N} }           { \partial \rho^{\upsilon} }                              
		\rho_N^s                          \sigma
		+ \dfrac{ \partial g_{\omega N} }            { \partial \rho^{\upsilon} } 
		\rho_N^\upsilon                     \omega_0
		+ \dfrac{ \partial g_{\rho N} }              { \partial \rho^{\upsilon} }          
		( \rho_n^\upsilon - \rho_p^\upsilon ) \rho_0  \\
		&+ \dfrac{ 1 }{ 2M_\Lambda } 
		\dfrac{ \partial f_{\omega\Lambda\Lambda} }{ \partial \rho^{\upsilon} } 
		\rho_\Lambda^T \omega_0.
	\end{split}
\end{equation}
The rearrangement term $\Sigma_R$ is present in the density-dependent CDFTs 
to ensure the energy-momentum conservation and thermodynamic consistency~\cite{Lenske1995_PLB345-355}.

In this work, the Bogoliubov transformation is implemented to describe pairing correlation between nucleons.  A separable pairing force of finite range in the spin-singlet channel~\cite{Tian2006_CPL23-3226,Tian2009_PLB676-44}, i.e.,
\begin{equation}
	V=-G\delta(R-R')P(r)P(r')\dfrac{1-P_\sigma}{2},	
\end{equation}
is adopted, in which $G$ the pairing strength, $R$ and $r$ the center of mass and relative coordinates, $P(r)$ the Gaussian function. The equal filling approximation~\cite{Perez-Martin2008_PRC78-014304} is adopted for the single $\Lambda$ hyperon. Then one can calculate the pairing field $\Delta$ and construct the relativistic Hartree Bogoliubov (RHB) equation
\begin{equation}
	\label{eq:rhb}
	\int d^{3}\bm{r}^{\prime}
	\left( \begin{array}{cc} h_B-\lambda  &  \Delta                      \\ 
		-\Delta^{*}   & -h_B+\lambda \end{array} 
	\right)
	\left( \begin{array}{c} U_{k} \\ V_{k} \end{array} \right)
	= E_{k}
	\left( \begin{array}{c} U_{k} \\ V_{k} \end{array} \right),
\end{equation}
where $\lambda$ is the Fermi energy, $E_k$ and $(U_k,V_k)^T$ are the quasi-particle energy and wave function, respectively. In MDC-CDFTs, substitute for treating the pairing with BCS approximation after solving Eq.~(\ref{Dirac_equ}), the method by solving Eq.~(\ref{eq:rhb}) directly is called the multidimensionally-constrained relativistic Hartree Bogoliubov (MDC-RHB) model.

The (hyper)nuclear shapes are charaterized by the deformation parameters $\beta_{\lambda\mu}$ as
\begin{equation}
	\beta_{\lambda\mu}=\dfrac{4\pi}{3AR^\lambda} Q_{\lambda\mu},
\end{equation}
where $R=1.2A^{1/3}$ fm with $A$ the mass number. $Q_{\lambda\mu}$ is the multipole moment of the intrinsic densities calculated as follow
\begin{equation}
	Q_{\lambda\mu}=\int d^3 r\rho^\upsilon(\bm r) r^\lambda Y_{\lambda\mu}(\Omega),
\end{equation}
where $Y_{\lambda\mu}(\Omega)$ is spherical harmonics with a Euler angle $\Omega=(\phi,\theta,\psi)$. 

In the MDC-RHB model, thanks to the usage of axial symmetric harmonic oscillator basis in solving the RHB equation~(\ref{eq:rhb}), one can keep or break the axial and reflection symmetries easily.
Therefore, four kinds of symmetries can be imposed: (a) axial-reflection symmetry with only $\beta_{20}$; (b) non-axial but reflection symmetry with $\beta_{20}$ and $\beta_{22}$; (c) axial symmetry but reflection asymmetry with $\beta_{20}$ and $\beta_{30}$; (d) non-axial and reflection asymmetry with $\beta_{20}$, $\beta_{22}$, $\beta_{30}$ and $\beta_{32}$, which are labeled as $K\pi$, $\slashed{K}\pi$, $K\slashed{\pi}$, and $\slashed{K}\slashed{\pi}$, respectively. 

Note that higher-order deformations are included naturely in MDC-CDFTs~\cite{Wang2022_CPC46-024107,Rong2023_PLB840-137896} with the $V_4$ symmetry, but we only discuss the deformations up to $\lambda=3$ in this work.

\section{RESULTS AND DISCUSSIONS}\label{sec:results}

\begin{figure}[htbp]
	\centering
	\includegraphics[width=0.48\textwidth]{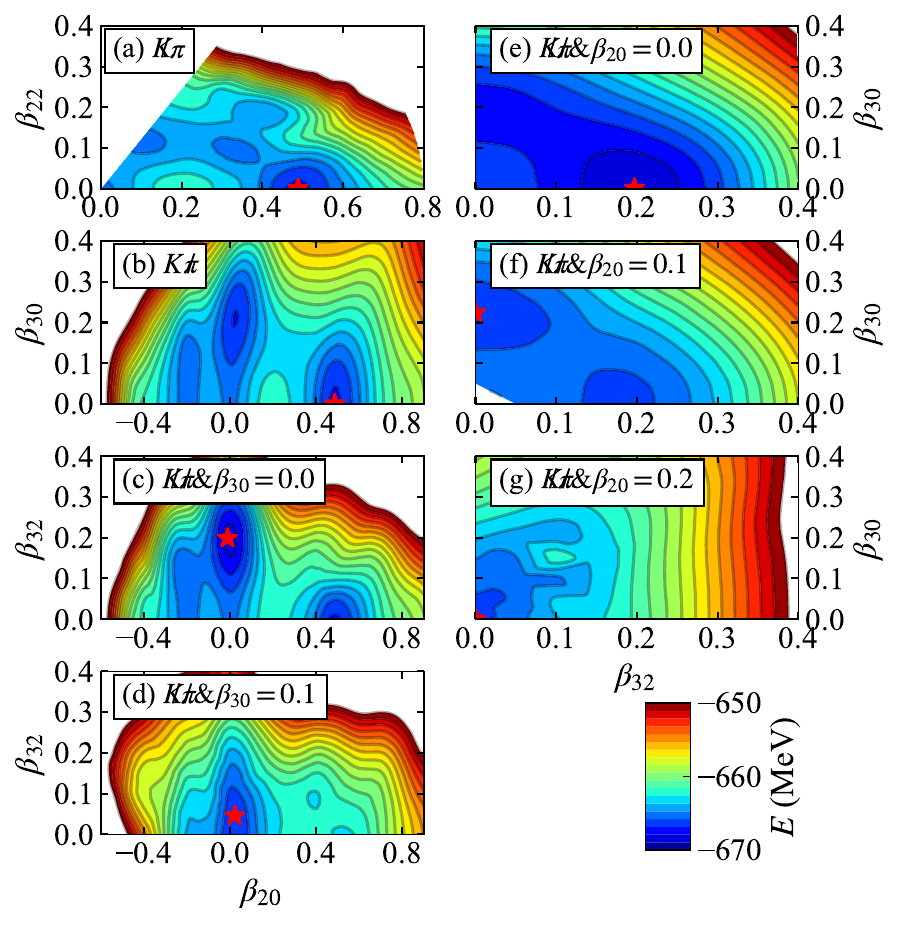}
	\caption{(Color online) The two-dimensional potential energy surfaces (PESs) with (a) $\slashed{K}\pi$, (b) $K\slashed{\pi}$, (c) $\slashed{K}\slashed{\pi}\&\beta_{30}=0.0$, (d) $\slashed{K}\slashed{\pi}\&\beta_{30}=0.1$, (e) $\slashed{K}\slashed{\pi}\&\beta_{20}=0.0$, (f) $\slashed{K}\slashed{\pi}\&\beta_{20}=0.1$, and (g) $\slashed{K}\slashed{\pi}\& \beta_{32}=0.2$ symmetry imposed. The PK1~\cite{Long2004_PRC69-034319} effective interaction is adopted for the particle-hole channel, and $G_n=728.00$ MeV fm$^3$ and $G_p=815.36$ MeV fm$^3$ are adopted for the pairing channel. The location of the global energy minimum on each PES is marked by red star.  }
	\label{fig:Zr80_PK1_2DPES_MF_Nf=14_Gp=1.12G0}
\end{figure}

To investigate the shape of $^{80}$Zr, we begin by calculating its two-dimensional PESs. Figure~\ref{fig:Zr80_PK1_2DPES_MF_Nf=14_Gp=1.12G0} shows the results obtained using the PK1~\cite{Long2004_PRC69-034319} effective interaction.
In these calculations, specific pairing strengths ($G_n=728.00$ MeV fm$^3$ and $G_p=815.36$ MeV fm$^3$) are assigned to the pairing forces between neutrons and protons within the nucleus, respectively. These parameters were adjusted to reproduce the available empirical pairing gaps of $^{102,104}$Zr~\cite{Zhao2017_PRC95-014320}.

In Fig.~\ref{fig:Zr80_PK1_2DPES_MF_Nf=14_Gp=1.12G0}(a), the Hill-Wheeler coordinates $\beta_2=\sqrt{\beta_{20}^2+2\beta_{22}^2}$ and $\gamma=\arctan(\sqrt{2}\beta_{22}/\beta_{20})$ with $\gamma \in [0,60^\circ]$ are constrained instead of $\beta_{20}$ and $\beta_{22}$ because of the six-fold symmetry with pure quadrupole deformations. The PES reveals the nucleus prefers a prolate shape in its ground state, with a quadrupole deformation of $\beta_2=0.49$. This finding aligns with several other theoretical models~\cite{Zheng2014_PRC90-064309,Lalazissis1999_AtDataNuclDataTables71-1,Zou2010_PRC82-024309,Moeller2016_ADNDT109-1} and experimental observations based on decay properties~\cite{Lister1987_PRL59-1270,Llewellyn2020_PRL124-152501}. However, it's noteworthy that some theoretical models predict a spherical ground state for $^{80}$Zr~\cite{Rodriguez2011_PLB705-255,Zhang2022_ADNDT144-101488}. Our calculations also show the presence of other possible shapes (spherical, oblate, and triaxial) at slightly higher energies on the PES, suggesting a potential coexistence of shapes. This finding is consistent with other studies~\cite{Rodriguez2011_PLB705-255,Zheng2014_PRC90-064309}.

Figure~\ref{fig:Zr80_PK1_2DPES_MF_Nf=14_Gp=1.12G0}(b) explores a different scenario where $K\slashed{\pi}$ symmetry is applied. In this case, the calculated PES reveals three possible low-energy shapes for the nucleus: prolate, oblate, and a new one called 'pear-like'. {\color{black}Interestingly, the pear-like shape, characterized by $(\beta_{20}, \beta_{30})=(0.02,0.20)$ deformation, has a slightly higher energy than the prolate shape.} This suggests that $^{80}$Zr naturally prefers a prolate shape. However, including the reflection asymmetry reduces the energy barrier between the prolate shape and the pear-like shape. This makes it easier for the nucleus to fluctuate between these two shapes, potentially even favoring the pear-like shape under certain conditions.

To explore a scenario with even less symmetry, we completely remove both reflection and axial restrictions (represented by $\slashed{K}\slashed{\pi}$ symmetry). In Fig.~\ref{fig:Zr80_PK1_2DPES_MF_Nf=14_Gp=1.12G0}(c), we fix $\beta_{30}=0.0$ and set $\beta_{22}$ as free parameter with initial value equals to zero, and calculate $E(\beta_{20}, \beta_{32})$. Similar to the previous case, the nucleus can adopt prolate and oblate shapes. However, a new, even lower-energy minimum appears, characterized by a deformation pattern of a tetrahedron. This finding aligns with calculations using other theoretical models~\cite{Tagami2014_PS89-054013,Tagami2015_JPG42-015106}.

We investigate how the interplay between various deformation parameters $(\beta_{20}, \beta_{30}, \beta_{32})$ affects the PESs of $^{80}$Zr. Figure~\ref{fig:Zr80_PK1_2DPES_MF_Nf=14_Gp=1.12G0}(d-g) present the results.
A key finding is that the tetrahedral minimum becomes more prominent when the $\beta_{30}$ and $\beta_{32}$ values are close to zero (Fig.~\ref{fig:Zr80_PK1_2DPES_MF_Nf=14_Gp=1.12G0}(c,e)). This suggests a more spherical shape favors the tetrahedral configuration. 
Additionally, the flatness of the PES with respect to $\beta_{30}$ and $\beta_{32}$ in Fig.~\ref{fig:Zr80_PK1_2DPES_MF_Nf=14_Gp=1.12G0}(e) indicates a high degree of shape fluctuation between the tetrahedral and pear-like shapes. As $\beta_{30}$ increases in $E(\beta_{20},\beta_{32})$ (Fig.~\ref{fig:Zr80_PK1_2DPES_MF_Nf=14_Gp=1.12G0}(d)), the correlation between the $Y_{20}$ and $Y_{32}$ weakens.
With increasing prolate deformation $\beta_{20}$ (Fig.~\ref{fig:Zr80_PK1_2DPES_MF_Nf=14_Gp=1.12G0}(f)), a saddle point emerges between the tetrahedral and axial-octupole energy minima. When $\beta_{20}$ reaches 0.2 (Fig.~\ref{fig:Zr80_PK1_2DPES_MF_Nf=14_Gp=1.12G0}(g)), both $\beta_{30}$ and $\beta_{32}$ do not impact the energy minimum of $^{80}$Zr.
These findings align with observations in neutron-rich Zr isotopes~\cite{Zhao2017_PRC95-014320} and support the notion that octupole deformations play a significant role in near-spherical nuclei~\cite{Takami1998_PLB431-242,Zberecki2006_PRC74-051302R,Zhao2017_PRC95-014320}. This further strengthens the validity of using the MDC-RHB model with $\slashed{K}\slashed{\pi}$ symmetry as a reliable approximation for studying the ground state of $^{80}$Zr.

\begin{figure}[htbp]
	\centering
	\includegraphics[width=0.48\textwidth]{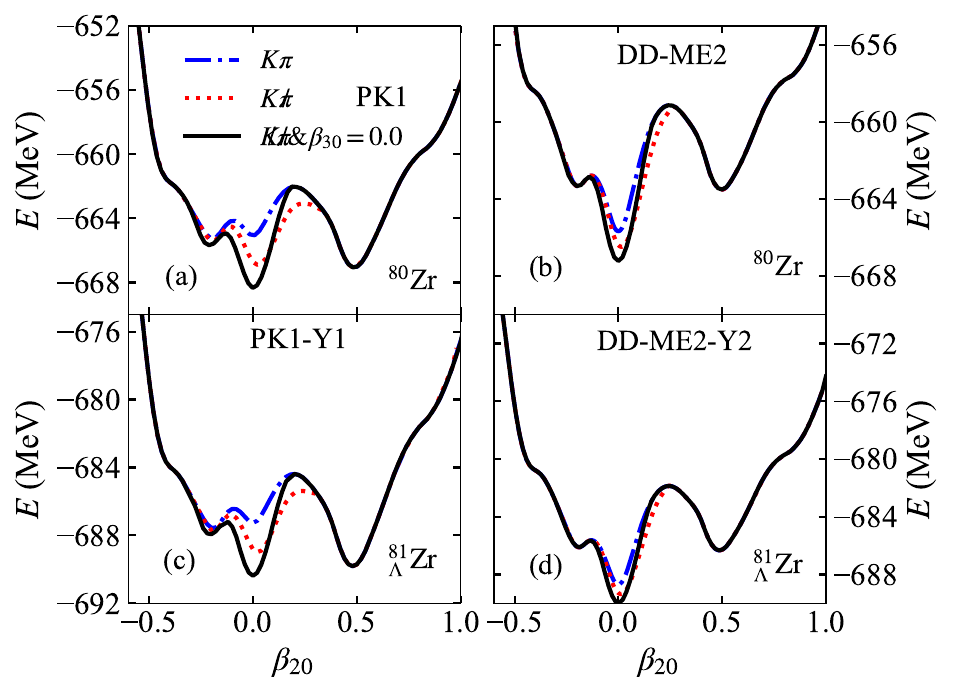}
	\caption{(Color online) The potential energy surfaces of (a) $^{80}$Zr with PK1, (b) $^{81}_\Lambda$Zr with PK1-Y1, (c) $^{80}$Zr with DD-ME2, and (d) $^{81}_\Lambda$Zr with DD-ME2-Y2. {The calculations are performed with $K\pi$ (blue dashed-dotted line), $K\slashed{\pi}$ (red dotted line), and $\slashed{K}\slashed{\pi} \&\beta_{30}=0.0$ (black solid line) symmetries imposed.}}
	\label{fig:Zr81L_PK1-Y1_DD-ME2-Y2}
\end{figure}

Figure~\ref{fig:Zr81L_PK1-Y1_DD-ME2-Y2} compares PESs calculated with various symmetry constraints in one-dimensional PESs. In Fig.~\ref{fig:Zr81L_PK1-Y1_DD-ME2-Y2}(a), the $K\pi$ symmetric PES exhibits three energy minima corresponding to those in Fig.~\ref{fig:Zr80_PK1_2DPES_MF_Nf=14_Gp=1.12G0}(a), except for the triaxial minimum. However, relaxing parity symmetry ($K\slashed{\pi}$ ) introduces an additional energy cost for near-spherical shapes. This is because a slightly non-spherical deformation term ($\beta_{30}$) becomes more influential in this case, favoring a pear-shaped minimum over a perfect sphere.
{In the $\slashed{K}\slashed{\pi}$ calculation, we fix $\beta_{30}$ value to zero to isolate the effect of $\beta_{30}$ and set the initial value of $\beta_{22}$ to zero.}
This leads to the spherical minimum transforming into a tetrahedral minimum. Interestingly, the oblate and prolate minima remain similar to the case with only $\beta_{20}$ considered. {It's worth noting that the higher energies observed with the $\slashed{K}\slashed{\pi} \& \beta_{30}=0.0$ symmetry than those with the $K\slashed{\pi} $ symmetry around the prolate region disappear when the $\beta_{30}$ and $\beta_{22}$ deformations are allowed to vary freely. We have checked this though the PESs in the supplement: A proper initial value of $\beta_{22}$ will make the energy of the prolate saddle point with $\slashed{K}\slashed{\pi}$ symmetry lower than the $K\slashed{\pi}$ one. Besides, the energy of the oblate minimum is lower by the $\beta_{22}$ deformation, but the ground state still exhibits tetrahedral shape. }

\begin{table}[htbp]
	\setlength\tabcolsep{11pt}
	\centering
	\caption{Minimums of $^{80}$Zr in the case of spherical, prolate, axial-octupole, and tetrahedral symmetry with different effective interactions using MDC-RHB model. The pairing strength $G_n=1.0 G_0$, $G_p=1.12 G_0$ with $G_0=728.0$ MeV fm$^3$ and effective range $a=0.644$ fm.}
	\begin{tabular}{ccccc}
		\hline
		\hline
		Interaction   & $\beta_{20}$  &  $\beta_{30}$ & $\beta_{32}$  &  $E$ (MeV) \\
		\hline
		\multirow{4}{*}{NL3}           &  0.000  & 0.000   &  0.000     & $-664.861$   \\
		{}            &  0.493    & 0.000 &  0.000     & $-665.191$    \\
		{}          &   0.015  &  0.185   & 0.000 & $-665.759$  \\
		{}            &  0.000    & 0.000 &  0.178   & $ -666.524 $   \\
		\hline
		\multirow{4}{*}{TM1}           &  0.000     &  0.000   & 0.000  & $-668.674$  \\
		{}            &  0.538  & 0.000   &  0.000     & $-664.996$  \\
		{}           &   0.017     &  0.194  &0.000   & $-669.800$  \\
		{}            &  0.000   & 0.000  &  0.184     & $-670.657$  \\
		\hline
		\multirow{4}{*}{PK1}           &  0.000     &  0.000   & 0.000  & $-665.074$  \\
		{}            &  0.491  & 0.000   &  0.000   & $-667.061$   \\
		{}            &  0.021     &  0.201   & 0.000  & $-666.894$   \\
		{}            &  0.000  & 0.000   &  0.192     & $-668.336$   \\
		\hline
		\multirow{4}{*}{PC-PK1}        &  0.000     &  0.000 & 0.000     & $-666.702$   \\
		{}            &  0.499  & 0.000   &  0.000     & $-663.911$   \\
		{}           &  0.005      & 0.118 & 0.000    & $-666.844$  \\
		{}            &  0.000  & 0.000   &  0.116     & $-666.971$   \\
		\hline
		\multirow{4}{*}{DD-ME2}        &  0.000     &  0.000 & 0.000     & $ -665.675$   \\
		{}            &   0.502  & 0.000   &  0.000     & $-663.500$   \\
		{}           &  0.012      &  0.166 & 0.000   & $-666.476$  \\
		{}            &  0.000  & 0.000   &  0.162     & $-667.213$   \\
		\hline

		\multirow{4}{*}{DD-PC1}        &  0.000     &  0.000 & 0.000    & $-667.551$   \\
		{}            &  0.536  & 0.000   &  0.000     & $-663.577$   \\
		{}            & 0.009      &  0.151 & 0.000    & $-668.032$  \\
		{}            &  0.000  & 0.000   &  0.144     & $-668.397$   \\
		\hline
		\hline
	\end{tabular}
	\label{tab:Min_interaction_Zr80}
\end{table}

Our calculations explore the dependence of ground state predictions on the effective interaction used. 
While the PK1 functional (Fig.~\ref{fig:Zr81L_PK1-Y1_DD-ME2-Y2}(a)) predicts prolate ground state with $K\pi$ symmetry, the DD-ME2~\cite{Lalazissis2005_PRC71-024312} interaction (Fig.~\ref{fig:Zr81L_PK1-Y1_DD-ME2-Y2}(b)) favors a spherical ground state with this symmetry. However, relaxing both axial and reflection symmetry leads to tetrahedral ground states for both effective interactions.
{To further solidify this result, we calculate the ground state of $^{80}$Zr using additional effective interactions (NL3~\cite{Lalazissis1997_PRC55-540}, TM1~\cite{Sugahara1994_NPA579-557}, PC-PK1~\cite{Zhao2010_PRC82-054319}, and DD-PC1~\cite{Niksic2008_PRC78-034318}) in Tab.~\ref{tab:Min_interaction_Zr80}.} Interestingly, all interactions predict a tetrahedral ground state for $^{80}$Zr. This consistency reinforces the conclusion that $^{80}$Zr exhibits a tetrahedral ground state, independent of the specific effective interaction employed.


Then, the PESs of $^{81}_\Lambda$Zr for various symmetries are studied using PK1-Y1~\cite{Wang2013_CTP60-479} (Fig.~\ref{fig:Zr81L_PK1-Y1_DD-ME2-Y2}(c)) and DD-ME2-Y2~\cite{Rong2021_PRC104-054321} (Fig.~\ref{fig:Zr81L_PK1-Y1_DD-ME2-Y2}(d)) effective interactions.
As expected, the PES for $^{81}_\Lambda$Zr with PK1-Y1 closely resembles that of its core nucleus, $^{80}$Zr (Fig.~\ref{fig:Zr81L_PK1-Y1_DD-ME2-Y2}(a)). 
This indicates that a single $\Lambda$ hyperon has minimal influence on the overall nuclear shape dominated by the eighty core nucleons. 
The inclusion of the $\beta_{30}$ (and $\beta_{32}$) deformation lowers the energies for near-spherical shapes, favoring a transition from a spherical minimum to a pear-like (or tetrahedral) minimum. 
Ultimately, the calculations predict a prolate minimum with slightly higher energy compared to the tetrahedral ground state. This suggests that $^{81}_\Lambda$Zr adopts a tetrahedral structure as the ground state, coexisting with prolate and pear-like shape isomers.
Similar results are observed with the DD-ME2-Y2 interaction (Fig.~\ref{fig:Zr81L_PK1-Y1_DD-ME2-Y2}(d)). Here, the overall PES is lowered upon binding a $\Lambda$ hyperon to $^{80}$Zr, but the core nucleons continue to dictate the dominant shape. The calculations again predict a tetrahedral ground state for $^{81}_\Lambda$Zr with a pear-like isomer.

\begin{figure*}[htbp]
	\centering
	\includegraphics[width=0.85\textwidth]{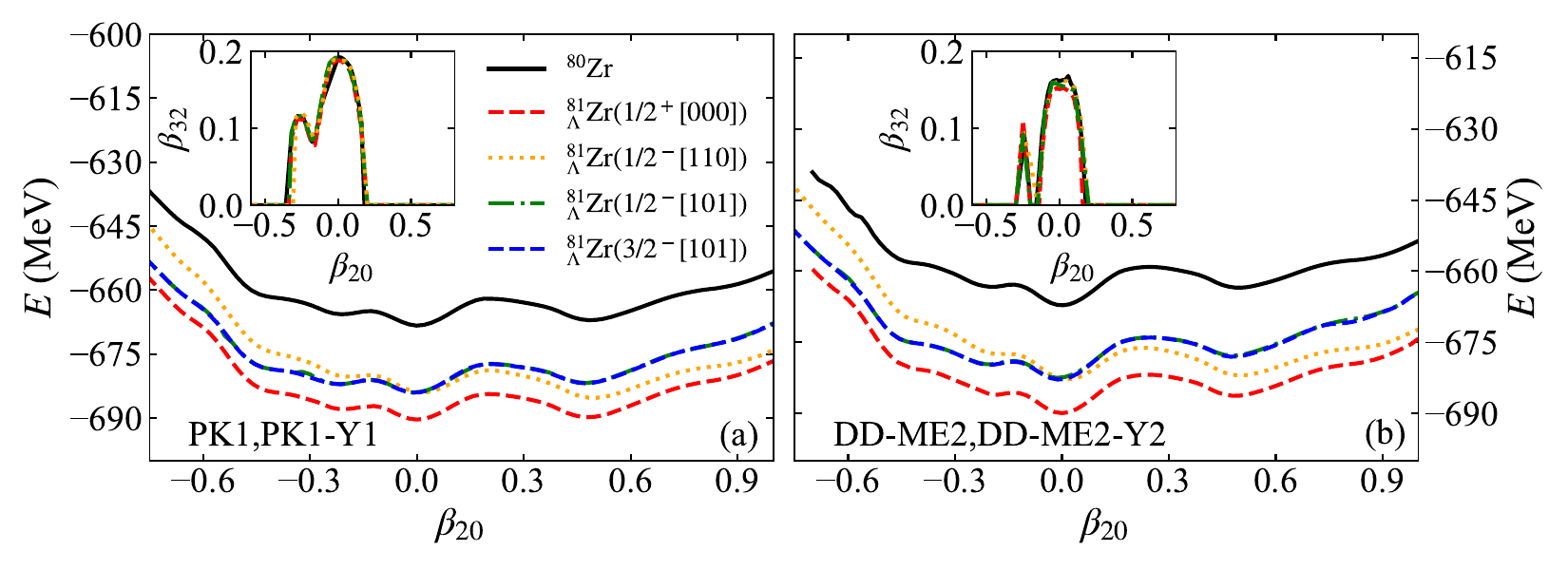}
	\caption{(Color online) The energies of $^{81}_\Lambda$Zr($^{80}$Zr) as a function of the deformation parameter $\beta_{20}$ calculated by (a) PK1-Y1(PK1) and (b) DD-ME2-Y2(DD-ME2) effective interactions. The calculations are performed with $\slashed{K}\slashed{\pi}\&\beta_{30}=0.0$ symmetry. In each subfigure, $\Lambda$ hyperon occupies the lowest $s$ orbit and the three $p$ orbits. The Nilsson quantum numbers $\Omega^\pi[N n_3 m_l]$ marked are the dominant component of the corresponding $\Lambda$ orbits. The inset in each subfigure shows the $\beta_{32}$ deformation of the corresponding PES as a function of $\beta_{20}$ deformation.}
	\label{fig:FF_PK1_Zr80&Zr81L}
\end{figure*}

\begin{table*}[htbp]
	\setlength\tabcolsep{11pt}
	\centering
	\caption{The deformation parameters ($\beta_{20},~\beta_{30}$ and $\beta_{32}$), root mean square (r.m.s.) matter radii ($r_m$), r.m.s. charge radii ($r_c$), total energies ($E$), $\Lambda$ separation energies $S_\Lambda$ of $^{81}_\Lambda$Zr and density overlaps between the nuclear core and hyperon ($I_{\rm overlap}$) with hyperon injected into the $\Lambda_s$ ($1/2^+[000]$) and $\Lambda_p$ ($1/2^-[110]$ and $1/2^-[101]$) orbits, respectively. The properties of $^{80}$Zr are also listed for comparison. The tetrahedral state (labeled with one asterisk) is calculated with $\slashed{K}\slashed{\pi}\& \beta_{30}=0.0$ symmetry, and the pear-like state (labeled with two asterisks) is calculated with $K\slashed{\pi}$ symmetry. }
	\label{tab:1} 
	\begin{tabular}{lcccccccc}
		\hline
		\hline
		(Hyper)nucleus & $\beta_{20}$  &  $\beta_{30}$  & $\beta_{32}$    &  $r_m$ (fm)  & $r_c$ (fm)  &  $E$ (MeV) & $S_\Lambda$ (MeV) & $I_{\rm overlap}$ (fm$^{-3}$)\\
		\hline
		\multicolumn{9}{c}{PK1-Y1~\cite{Wang2013_CTP60-479}} \\
		$^{80}$Zr &  0.491       &  --      &  -- & 4.219  &  4.326   & $-667.019$  &   --     &  --    \\
		$^{81}_\Lambda$Zr($1/2^+[000]$) 
		&  0.483    &  --     &  --      & 4.202  &  4.320   & $-689.764$  & 22.745  & 0.144 \\
		$^{81}_\Lambda$Zr($1/2^-[110]$) 
		&  0.498    &  --     &  --      & 4.218  &  4.327   & $-685.247$  & 18.228  & 0.139 \\
		$^{81}_\Lambda$Zr($1/2^-[101]$) 
		&  0.475    &  --     &  --      & 4.208  & 4.321    & $-681.709$  & 14.690  & 0.124   \\
		$^{80}$Zr$^*$ 
		&  0.000    &  0.000     &   0.192    & {4.154}  &  4.265   & $-668.332$  &    --     & --     \\
		$^{81}_\Lambda$Zr$^*$($1/2^+[000]$) 
		&  0.000    &  0.000     &   0.189    & 4.140  &  4.263   &{$-690.372$}  & {22.040}  & 0.142  \\
		$^{81}_\Lambda$Zr$^*$($1/2^-[110]$) 
		&  0.011    &  0.000     &   0.192    & 4.151  &  4.265   & $-684.078$  & 15.746  & 0.126 \\
		$^{81}_\Lambda$Zr$^*$($1/2^-[101]$)  
		&  0.000    &  0.000     &   0.192    & 4.151  & 4.265    & $-683.957$  & 15.625  &  0.126  \\
		$^{80}$Zr$^{**}$ 
		&  0.021    & 0.202       &  --     & 4.139  &  4.252  & $-666.895$  &    --    &  --    \\
		$^{81}_\Lambda$Zr$^{**}$($1/2^+[000]$) 
		&  0.020    & 0.197       &  --     & 4.126  &  4.250  & $-688.984$  & 22.089  &  0.143 \\
		$^{81}_\Lambda$Zr$^{**}$($1/2^-[110]$) 
		&  0.037    & 0.206       &  --     & 4.137  &  4.253  & $-682.767$  & 15.872  &  0.127  \\
		$^{81}_\Lambda$Zr$^{**}$($1/2^-[101]$) 
		&  0.015    & 0.199       &  --     & 4.136  &  4.252  & $-682.432$  & 15.537  &  0.125 \\
		\hline
		\multicolumn{9}{c}{DD-ME2-Y2~\cite{Rong2021_PRC104-054321}} \\
		$^{80}$Zr &  0.502    &  --      &  --     & 4.240   & 4.350  & $-663.500$  &   --     &  --  \\
		$^{81}_\Lambda$Zr($1/2^+[000]$) 
		&  0.491    &  --      &  --     & 4.206   & 4.329  & $-686.322$  & 22.822  &  0.149 \\
		$^{81}_\Lambda$Zr($1/2^-[110]$) 
		&  0.506    &  --      &  --     & 4.224   & 4.337  & $-682.011$  & 18.511  &  0.146 \\
		$^{81}_\Lambda$Zr($1/2^-[101]$) 
		&  0.483    &  --     &  --     & 4.214   & 4.331  & $-677.811$  & 14.311  &  0.130  \\
		$^{80}$Zr$^*$ 
		&  0.000    &  0.000      &  0.162     & 4.157   & 4.272  & $-667.213$  &  --      &  -- \\
		$^{81}_\Lambda$Zr$^*$($1/2^+[000]$) 
		&  0.000    &  0.000      &  0.151     & 4.125   & 4.254  & $-689.973$  &  22.760 &  0.152 \\
		$^{81}_\Lambda$Zr$^*$($1/2^-[110]$) 
		& {0.013}    &  0.000      &  0.161     & 4.140   & 4.261  & {$-682.948$}  &  {15.735} &  0.135 \\
		$^{81}_\Lambda$Zr$^*$($1/2^-[101]$) 
		&  {0.000}    &  0.000      &  {0.160}     & 4.140   & 4.261  & {$-682.762$}  &  {15.549} &  0.136  \\
		$^{80}$Zr$^{**}$ 
		&  0.012    & {0.166}      &  --     & 4.147   & 4.264  & {$-666.476$}  &  --      &  --    \\
		$^{81}_\Lambda$Zr$^{**}$($1/2^+[000]$) 
		&  0.010    &  0.154      &  --     & 4.117   & 4.247  & $-689.352$  & {22.876}  &  0.153    \\
		$^{81}_\Lambda$Zr$^{**}$($1/2^-[110]$) 
		&  0.027    &  0.177      &  --     & 4.132   & 4.254  & $-682.403$  & {15.927}  &  0.137   \\
		$^{81}_\Lambda$Zr$^{**}$($1/2^-[101]$) 
		&  0.006    &  0.161      &  --     & 4.130   & 4.252  & $-681.940$  & {15.464}  &  0.134   \\
		\hline
		\hline		
	\end{tabular}
\end{table*}

Therefore, to investigate the effects of the $\Lambda$ hyperon impurity in $^{81}_\Lambda$Zr, calculations that relax both axial and reflection symmetry constraints are necessary (denoted as $\slashed{K}\slashed{\pi}$). 
For simplicity, Fig.~\ref{fig:FF_PK1_Zr80&Zr81L} presents results from calculations with $\slashed{K}\slashed{\pi} \& \beta_{30} = 0.0$. As discussed previously, $\beta_{30}$ influences the PES within the $-0.2<\beta_{20}<0.2$ region but doesn't alter the energy minima.
In this investigation, we calculate the PESs for $^{81}_\Lambda$Zr where the $\Lambda$ hyperon occupies the four lowest available orbits. 
These PESs are then compared to the PES of the core nucleus, $^{80}$Zr.
It's important to note that while the Nilsson quantum numbers ($\Omega^\pi[N n_3 m_l]$) are used to label the $\Lambda$ single-particle states for convenience, these lose their exact meaning when both axial and reflection symmetries are broken.

Figure~\ref{fig:FF_PK1_Zr80&Zr81L} reveals a strong similarity between the PESs of $^{81}_\Lambda$Zr and those of its core nucleus, $^{80}$Zr, for both PK1-Y1 and DD-ME2-Y2 interactions. {The presence of a $\Lambda$ hyperon occupying the $1/2^+[000]$ orbit primarily leads to an binding energy increase of around 22 MeV.} Interestingly, the $1/2^-[101]$ and $3/2^-
[101]$ states exhibit nearly identical energies for specific shapes. This is naturally attributed to the very weak spin-orbit splitting observed in $\Lambda$ hyperons, as reported in Ref.~\cite{Ajimura2001_PRL86-4255}.
The inset of the figure depicts the $\beta_{32}$ deformation of the PESs as a function of the $\beta_{20}$ deformation. Notably, for all PESs, the $\beta_{32}$ values coincide with those of $^{80}$Zr for specific $\beta_{20}$ values.



\begin{figure*}[htbp]
	\centering
	\includegraphics[width=0.65\textwidth]{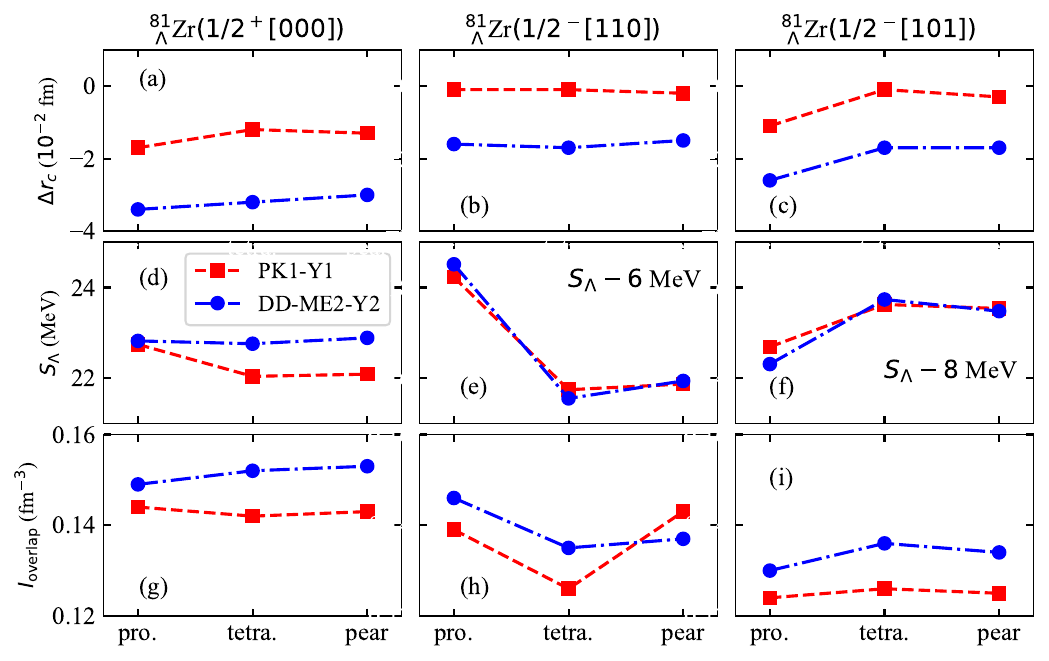}
	\caption{(Color online) The difference of the core and hypernuclear charge radii $\Delta r_c$, $\Lambda$ separation energy ($S_\Lambda$), and density overlap of the core and hyperon $I_{\rm overlap}$ as a function of the isomer (prolate, tetrahedral and pear-like) states. The $\Lambda$ hyperon occupies the lowest $s$ orbit ($\Omega^\pi[Nn_3m_l]=1/2^+[000]$) and the two $p$ orbits ($1/2^-[110]$ and $1/2^-[101]$). For the convenience of comparison, panels (e) and (f) are offset downward from panel (d) by $6$ and $8$ MeV, respectively. The calculations are done by using PK1-Y1 and DD-ME2-Y2 effective interactions, respectively. }
	\label{fig:Zr81L_SL_IO_Rm}
\end{figure*}

Table~\ref{tab:1} provides quantitative data to explore the relationship between nuclear shape and the influence of the $\Lambda$ hyperon impurity. Let's focus on the deformation changes. When a $\Lambda$ particle occupies the $1/2^+
[000]$ state, all three studied states (prolate, pear-like, and tetrahedral) exhibit a decrease in deformation compared to their counterparts in the core nucleus.
{For the $1/2^-[110] (1/2^-
[101])$ state occupancy by a hyperon, the $\beta_{20}$ deformation of the prolate state increases, while the $\beta_{30}$ deformation of the pear-like state decreases compared to the core nucleus. Interestingly, the $\beta_{32}$ deformation of the tetrahedral state remains nearly unchanged.}

In addition to the deformation changes, our findings also reveal a shrinkage effect. The prolate state exhibits a larger matter radius and charge radius compared to the tetrahedral and pear-like states. 
This is further illustrated in Fig.~\ref{fig:Zr81L_SL_IO_Rm}, which depicts the difference in core and hypernuclear charge radii ($\Delta r_c$), defined as $\Delta r_c = r_c(^{81}_{~\Lambda} {\rm Zr}) - r_c(^{80}{\rm Zr})$. Here, $r_c$ signifies the charge radius. 
The $\Delta r_c$ values range from $-0.04$ fm to 0.01 fm when the $\Lambda$ particle occupies the $s_{1/2}$ orbit. This shrinkage effect is smaller for $p$-orbit occupancy and demonstrates sensitivity to the chosen effective interaction. Calculations using the PK1-Y1 interaction predict a smaller shrinkage compared to those using DD-ME2-Y2.

The final aspect we explore is the $\Lambda$ separation energy, denoted by $S_\Lambda$, which represents the energy difference between the hypernucleus ($^{81}_{~\Lambda}{\rm Zr}$) and the core nucleus ($^{80}$Zr) expressed as: $S_\Lambda = E({}^{81}_\Lambda {\rm Zr})-E(^{80}{\rm Zr})$. 
Figure~\ref{fig:Zr81L_SL_IO_Rm}(d) depicts this energy for the three studied states when a $\Lambda$ particle occupies the $s_{1/2}$ orbit. The $S_\Lambda$ values are around 22-23 MeV, with minor variations between the shapes. 
Notably, the prolate state exhibits a slightly higher separation energy compared to the other two states when calculated with the PK1-Y1 interaction. In contrast, the DD-ME2-Y2 interaction predicts the highest separation energy for the axial-octupole state. Anyway, the tetrahedral ground state does not exhibit the strongest binding for the $\Lambda$ hyperon, regardless of the interaction.
For the $\Lambda$ particle occupying the $1/2^-[110] $ orbit, the $S_\Lambda$ value for the prolate state is approximately 2 MeV {larger than} the pear-like and tetrahedral states. Interestingly, when $\Lambda$ occupies $1/2^-[101]$, the largest separation energy is consistently found in the tetrahedral shape, irrespective of the chosen effective interaction.

\begin{figure*}[htbp]
	\centering
	\includegraphics[width=0.75\textwidth]{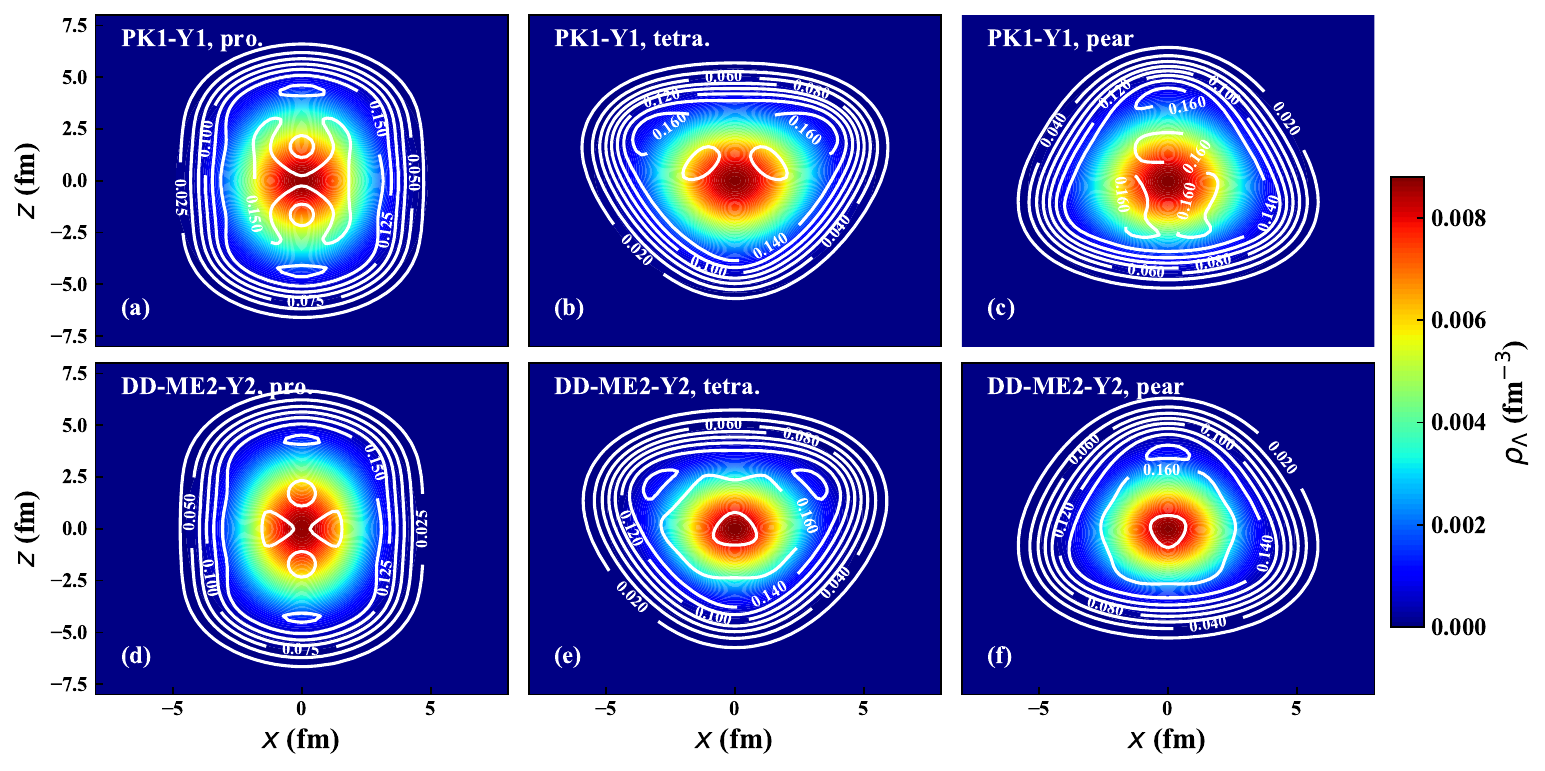}
	\caption{(Color online) The contour plot of nucleon density and the density profile of the $\Lambda$ hyperon in the $(x,z)$ plane at $y=0$ fm for differrent shapes of $^{81}_{\Lambda}$Zr calculated with PK1-Y1 (a-c) and DD-ME2-Y2 (d-f) effective interactions, respectively. The $\Lambda$ hyperon occupies the $1/2^+[000]$ orbit.}
	\label{fig:Zr81L_12000_dens_Nf=14_1.0Gn_1.12Gp}
\end{figure*}

Finally, the overlap integral ($I_{\rm overlap}$) between the core nucleon density ($\rho_N(r)$) and the $\Lambda$ density ($\rho_\Lambda(r)$) is calculated using the equation~\cite{Win2011_PRC83-014301,Lu2014_PRC89-044307,Zhou2016_PRC94-024331,Wu2017_PRC95-034309}: 
\begin{equation}
	I_{\rm overlap} = \int \rho_N(r)\rho_\Lambda(r) d^3r. 
\end{equation}
This integral quantifies the degree of spatial overlap between these densities. {It was introduced in Ref.~\cite{Win2011_PRC83-014301}. One found slightly softer energy curve toward the prolate configuration in C isotopes originates from the fact that the overlap between the deformed nuclear density and a spherical $\Lambda$ density is maximum at the prolate configuration. Later, a correlation between $S_\Lambda$ and $I_{\rm overlap}$ is found is Refs.~\cite{Lu2014_PRC89-044307,Zhou2016_PRC94-024331,Wu2017_PRC95-034309}.} In this work,
As observed in Table~\ref{tab:1} and Fig.~\ref{fig:Zr81L_SL_IO_Rm}(d-i), a larger $\Lambda$ separation energy ($S_\Lambda$) generally correlates with a larger overlap integral. However, an exception occurs for the case of the $\Lambda$ particle occupying the $1/2^+[000]$ state with the DD-ME2-Y2 interaction. Here, the prolate state exhibits a higher $S_\Lambda$ than the pear-like state, despite a lower $I_{\rm overlap}$ value. This discrepancy can be understood by examining the nucleon and $\Lambda$ density distributions, which are presented in Fig.~\ref{fig:Zr81L_12000_dens_Nf=14_1.0Gn_1.12Gp} for calculations with both PK1-Y1 and DD-ME2-Y2 interactions. In all cases, the $\Lambda$ hyperon resides in the center of the $x-z$ plane. However, the tetrahedral state exhibits a "clustered" feature within its nucleon density, leading to a lower density at the center of the $x-z$ plane compared to the other shapes. This explains the smaller $S_\Lambda$ observed for the tetrahedral state. Conversely, the large $S_\Lambda$ value for the pear-like state with DD-ME2-Y2 interaction arises from its very small radii and a more centralized nucleon density distribution.

\section{SUMMARY}\label{sec:summary}

This study investigates the $Y_{32}$ correlation in $^{80}$Zr and Lambda impurity effect in tetrahedral shape using the MDC-RHB model. We find that the ground state of $^{80}$Zr exhibits a minimum energy configuration in a tetrahedral shape, along with minima in prolate and axial-octupole shapes. The octupole deformations $\beta_{30}$ and $\beta_{32}$ lower the PESs in near-spherical shapes. However, differentiating between the pear-like and tetrahedral shapes remains challenging due to the very flat potential energy surface $E(\beta_{30}, \beta_{32})$ in near-spherical region.

The binding strength of the $\Lambda$ particle, quantified by the $\Lambda$ separation energy ($S_\Lambda$), depends on the specific energy level (orbital) it occupies within the nucleus and the chosen nuclear interaction. For the $\Lambda$ in the $1/2^+[000]$ and $1/2^-[110]$ state, the prolate state shows the largest $S_\Lambda$ with the PK1-Y1 interaction, while the DD-ME2-Y2 interaction favors the axial-octupole state. Interestingly, when the $\Lambda$ occupies the $1/2^-[101]$ orbit, the tetrahedral shape consistently exhibits the strongest binding (largest $S_\Lambda$) regardless of the interaction used.
Our analysis reveals a general correlation between $S_\Lambda$ and the overlap integral ($I_{\rm overlap}$) between the Lambda and nucleon densities. This suggests a stronger binding for a larger spatial overlap between these densities. However, an exception occurs for the $1/2^+[000]$ state with the DD-ME2-Y2 interaction. In this case, a "clustered" structure within the tetrahedral nucleon density distribution leads to a smaller $S_\Lambda$ despite a potentially larger overlap.


	\begin{acknowledgments}
		This work has been supported by the National Natural Science Foundation of China (Grant No. 12205057), the Science and Technology Plan Project of Guangxi (Grants No. Guike AD23026250),
		the Central Government Guides Local Scientific and Technological Development Fund Projects (Grants No. Guike ZY22096024), the Research Basic Ability Enhancement
		Project for Young and Middle-aged Teachers in the Universities in Guangxi (Grant No. 2022KY0055), the National Natural Science Foundation of China (Grant No. 12365016), and the Natural
		Science Foundation of Guangxi (Grant No. 2023GXNSFAA026016).
	\end{acknowledgments}

    \section*{Appendix}
\begin{figure}[htbp]
		\centering
		\includegraphics[width=0.48\textwidth]{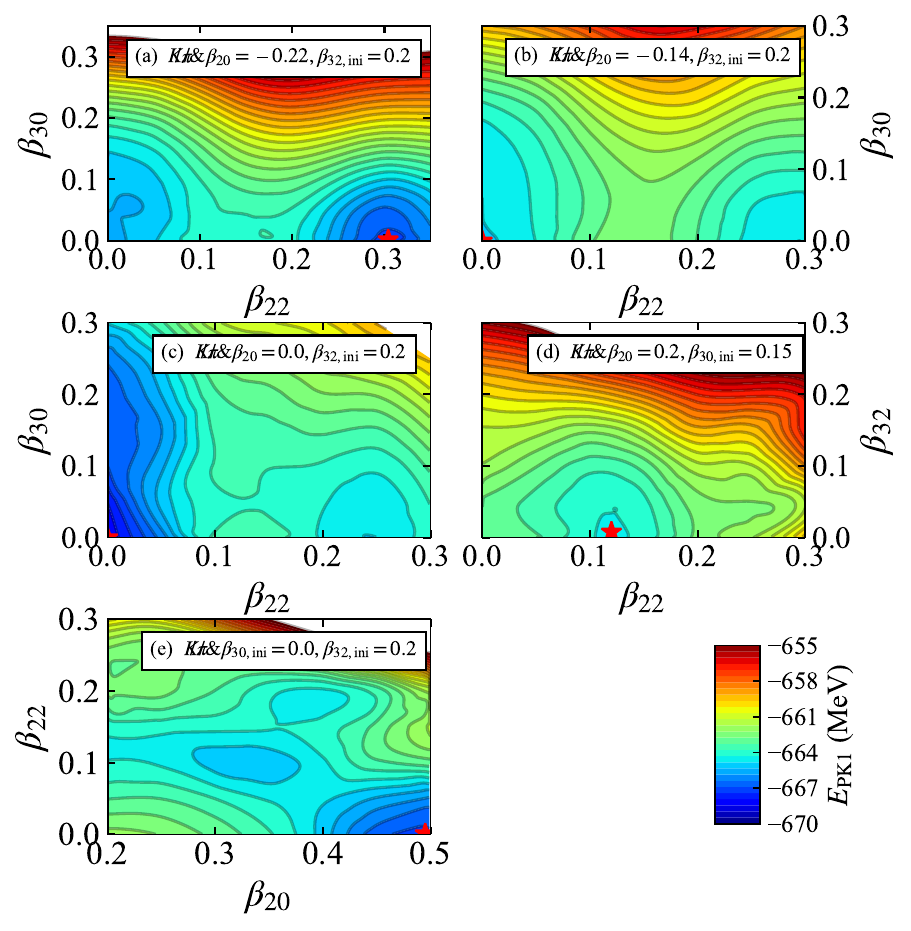}
		\caption{(Color online) The two-dimensional potential energy surfaces (PESs) at (a) the $E(\beta_{20}=-0.22,\beta_{32})$ energy minimum, (b) the $E(\beta_{20}=-0.14,\beta_{32})$ saddle point, (c) the tetrahedral energy minimum, (d) the $E(\beta_{20}=0.2,\beta_{30})$ saddle point, and (e) the prolate region with $\beta_{20}\textgreater 0.2$ of Fig.~\ref{fig:Zr81L_PK1-Y1_DD-ME2-Y2}(a). The location of the global energy minimum on each PES is marked by red star.}
		\label{fig:supplement_80Zr_2DPES}
	\end{figure}

    	The 2DPESs of $^{80}$Zr at different points of Fig.~\ref{fig:Zr81L_PK1-Y1_DD-ME2-Y2}(a). Fig.~\ref{fig:supplement_80Zr_2DPES}(a) is the $E(\beta_{22},\beta_{30})$ from the energy minimum of $E(\beta_{20}=-0.22,\beta_{32})$. The energy minimum appears at $\beta_{22}\approx 0.3$ and $\beta_{30}=0.0$, meaning the triaxial shape will lower the energy of the energy minimum of $E(\beta_{20}=-0.22,\beta_{32})$. However, the obtained energy is still higher than the tetrahedral one. Fig.~\ref{fig:supplement_80Zr_2DPES}(b) and (c) are the $E(\beta_{22},\beta_{30})$ from the saddle point of $E(\beta_{20}=-0.14,\beta_{32})$ and the tetrahedral minimum, respectively. The energy minima appear at $\beta_{22}=0.0$ and $\beta_{30}=0.0$, indicating these two deformation parameters do not attribute to the energies at the $E(\beta_{20}=-0.14,\beta_{32})$ saddle point and the tetrahedral minimum. Fig.~\ref{fig:supplement_80Zr_2DPES}(d) is the $E(\beta_{22},\beta_{32})$ from the saddle point of $E(\beta_{20}=0.2,\beta_{30})$. The energy minimum appears at $\beta_{22}\approx 0.12$ and $\beta_{32}\approx0.0$ and lower than the $K\slashed{\pi}$ saddle point at $\beta_{20}=0.2$. Fig.~\ref{fig:supplement_80Zr_2DPES}(e) shows the PES with $\beta_{20}$ from 0.2 to 0.5. Combining the obtained result that octupole deformations will not change the energies in this region, this figure tells us the energy minimum at $\beta_{20}\approx 0.5$ is prolate in the $\slashed{K}\slashed{\pi}$ calculations. 

\bibliographystyle{apsrev4-2}
	\bibliography{../../../Notes/bib/ref}

\begin{thebibliography}{86}%
\makeatletter
\providecommand \@ifxundefined [1]{%
 \@ifx{#1\undefined}
}%
\providecommand \@ifnum [1]{%
 \ifnum #1\expandafter \@firstoftwo
 \else \expandafter \@secondoftwo
 \fi
}%
\providecommand \@ifx [1]{%
 \ifx #1\expandafter \@firstoftwo
 \else \expandafter \@secondoftwo
 \fi
}%
\providecommand \natexlab [1]{#1}%
\providecommand \enquote  [1]{``#1''}%
\providecommand \bibnamefont  [1]{#1}%
\providecommand \bibfnamefont [1]{#1}%
\providecommand \citenamefont [1]{#1}%
\providecommand \href@noop [0]{\@secondoftwo}%
\providecommand \href [0]{\begingroup \@sanitize@url \@href}%
\providecommand \@href[1]{\@@startlink{#1}\@@href}%
\providecommand \@@href[1]{\endgroup#1\@@endlink}%
\providecommand \@sanitize@url [0]{\catcode `\\12\catcode `\$12\catcode
  `\&12\catcode `\#12\catcode `\^12\catcode `\_12\catcode `\%12\relax}%
\providecommand \@@startlink[1]{}%
\providecommand \@@endlink[0]{}%
\providecommand \url  [0]{\begingroup\@sanitize@url \@url }%
\providecommand \@url [1]{\endgroup\@href {#1}{\urlprefix }}%
\providecommand \urlprefix  [0]{URL }%
\providecommand \Eprint [0]{\href }%
\providecommand \doibase [0]{https://doi.org/}%
\providecommand \selectlanguage [0]{\@gobble}%
\providecommand \bibinfo  [0]{\@secondoftwo}%
\providecommand \bibfield  [0]{\@secondoftwo}%
\providecommand \translation [1]{[#1]}%
\providecommand \BibitemOpen [0]{}%
\providecommand \bibitemStop [0]{}%
\providecommand \bibitemNoStop [0]{.\EOS\space}%
\providecommand \EOS [0]{\spacefactor3000\relax}%
\providecommand \BibitemShut  [1]{\csname bibitem#1\endcsname}%
\let\auto@bib@innerbib\@empty
\bibitem [{\citenamefont {Dudek}\ \emph {et~al.}(2002)\citenamefont {Dudek},
  \citenamefont {Go\ifmmode \acute{z}\else \'{z}\fi{}d\ifmmode~\acute{z}\else
  \'{z}\fi{}}, \citenamefont {Schunck},\ and\ \citenamefont
  {Mi\ifmmode~\acute{s}\else \'{s}\fi{}kiewicz}}]{Dudek2002_PRL88-252502}%
  \BibitemOpen
  \bibfield  {author} {\bibinfo {author} {\bibfnamefont {J.}~\bibnamefont
  {Dudek}}, \bibinfo {author} {\bibfnamefont {A.}~\bibnamefont {Go\ifmmode
  \acute{z}\else \'{z}\fi{}d\ifmmode~\acute{z}\else \'{z}\fi{}}}, \bibinfo
  {author} {\bibfnamefont {N.}~\bibnamefont {Schunck}},\ and\ \bibinfo {author}
  {\bibfnamefont {M.}~\bibnamefont {Mi\ifmmode~\acute{s}\else
  \'{s}\fi{}kiewicz}},\ }\href {https://doi.org/10.1103/PhysRevLett.88.252502}
  {\bibfield  {journal} {\bibinfo  {journal} {Phys. Rev. Lett.}\ }\textbf
  {\bibinfo {volume} {88}},\ \bibinfo {pages} {252502} (\bibinfo {year}
  {2002})}\BibitemShut {NoStop}%
\bibitem [{\citenamefont {Takami}\ \emph {et~al.}(1998)\citenamefont {Takami},
  \citenamefont {Yabana},\ and\ \citenamefont
  {Matsuo}}]{Takami1998_PLB431-242}%
  \BibitemOpen
  \bibfield  {author} {\bibinfo {author} {\bibfnamefont {S.}~\bibnamefont
  {Takami}}, \bibinfo {author} {\bibfnamefont {K.}~\bibnamefont {Yabana}},\
  and\ \bibinfo {author} {\bibfnamefont {M.}~\bibnamefont {Matsuo}},\ }\href
  {https://doi.org/10.1016/S0370-2693(98)00545-0} {\bibfield  {journal}
  {\bibinfo  {journal} {Phys. Lett. B}\ }\textbf {\bibinfo {volume} {431}},\
  \bibinfo {pages} {242} (\bibinfo {year} {1998})}\BibitemShut {NoStop}%
\bibitem [{\citenamefont {Yamagami}\ \emph {et~al.}(2001)\citenamefont
  {Yamagami}, \citenamefont {Matsuyanagi},\ and\ \citenamefont
  {Matsuo}}]{Yamagami2001_NPA693-579}%
  \BibitemOpen
  \bibfield  {author} {\bibinfo {author} {\bibfnamefont {M.}~\bibnamefont
  {Yamagami}}, \bibinfo {author} {\bibfnamefont {K.}~\bibnamefont
  {Matsuyanagi}},\ and\ \bibinfo {author} {\bibfnamefont {M.}~\bibnamefont
  {Matsuo}},\ }\href
  {https://doi.org/https://doi.org/10.1016/S0375-9474(01)00918-6} {\bibfield
  {journal} {\bibinfo  {journal} {Nucl. Phys. A}\ }\textbf {\bibinfo {volume}
  {693}},\ \bibinfo {pages} {579} (\bibinfo {year} {2001})}\BibitemShut
  {NoStop}%
\bibitem [{\citenamefont {Tagami}\ \emph {et~al.}(2014)\citenamefont {Tagami},
  \citenamefont {Shimada}, \citenamefont {Fujioka}, \citenamefont {Shimizu},\
  and\ \citenamefont {Dudek}}]{Tagami2014_PS89-054013}%
  \BibitemOpen
  \bibfield  {author} {\bibinfo {author} {\bibfnamefont {S.}~\bibnamefont
  {Tagami}}, \bibinfo {author} {\bibfnamefont {M.}~\bibnamefont {Shimada}},
  \bibinfo {author} {\bibfnamefont {Y.}~\bibnamefont {Fujioka}}, \bibinfo
  {author} {\bibfnamefont {Y.~R.}\ \bibnamefont {Shimizu}},\ and\ \bibinfo
  {author} {\bibfnamefont {J.}~\bibnamefont {Dudek}},\ }\href
  {https://doi.org/10.1088/0031-8949/89/5/054013} {\bibfield  {journal}
  {\bibinfo  {journal} {Phys. Scr.}\ }\textbf {\bibinfo {volume} {89}},\
  \bibinfo {pages} {054013} (\bibinfo {year} {2014})}\BibitemShut {NoStop}%
\bibitem [{\citenamefont {Tagami}\ \emph {et~al.}(2015)\citenamefont {Tagami},
  \citenamefont {Shimizu},\ and\ \citenamefont
  {Dudek}}]{Tagami2015_JPG42-015106}%
  \BibitemOpen
  \bibfield  {author} {\bibinfo {author} {\bibfnamefont {S.}~\bibnamefont
  {Tagami}}, \bibinfo {author} {\bibfnamefont {Y.~R.}\ \bibnamefont
  {Shimizu}},\ and\ \bibinfo {author} {\bibfnamefont {J.}~\bibnamefont
  {Dudek}},\ }\href {https://doi.org/10.1088/0954-3899/42/1/015106} {\bibfield
  {journal} {\bibinfo  {journal} {J. Phys. G: Nucl. Part. Phys.}\ }\textbf
  {\bibinfo {volume} {42}},\ \bibinfo {pages} {015106} (\bibinfo {year}
  {2015})}\BibitemShut {NoStop}%
\bibitem [{\citenamefont {Lister}\ \emph {et~al.}(1987)\citenamefont {Lister},
  \citenamefont {Campbell}, \citenamefont {Chishti}, \citenamefont {Gelletly},
  \citenamefont {Goettig}, \citenamefont {Moscrop}, \citenamefont {Varley},
  \citenamefont {James}, \citenamefont {Morrison}, \citenamefont {Price},
  \citenamefont {Simpson}, \citenamefont {Connel},\ and\ \citenamefont
  {Skeppstedt}}]{Lister1987_PRL59-1270}%
  \BibitemOpen
  \bibfield  {author} {\bibinfo {author} {\bibfnamefont {C.~J.}\ \bibnamefont
  {Lister}}, \bibinfo {author} {\bibfnamefont {M.}~\bibnamefont {Campbell}},
  \bibinfo {author} {\bibfnamefont {A.~A.}\ \bibnamefont {Chishti}}, \bibinfo
  {author} {\bibfnamefont {W.}~\bibnamefont {Gelletly}}, \bibinfo {author}
  {\bibfnamefont {L.}~\bibnamefont {Goettig}}, \bibinfo {author} {\bibfnamefont
  {R.}~\bibnamefont {Moscrop}}, \bibinfo {author} {\bibfnamefont {B.~J.}\
  \bibnamefont {Varley}}, \bibinfo {author} {\bibfnamefont {A.~N.}\
  \bibnamefont {James}}, \bibinfo {author} {\bibfnamefont {T.}~\bibnamefont
  {Morrison}}, \bibinfo {author} {\bibfnamefont {H.~G.}\ \bibnamefont {Price}},
  \bibinfo {author} {\bibfnamefont {J.}~\bibnamefont {Simpson}}, \bibinfo
  {author} {\bibfnamefont {K.}~\bibnamefont {Connel}},\ and\ \bibinfo {author}
  {\bibfnamefont {O.}~\bibnamefont {Skeppstedt}},\ }\href
  {https://doi.org/10.1103/PhysRevLett.59.1270} {\bibfield  {journal} {\bibinfo
   {journal} {Phys. Rev. Lett.}\ }\textbf {\bibinfo {volume} {59}},\ \bibinfo
  {pages} {1270} (\bibinfo {year} {1987})}\BibitemShut {NoStop}%
\bibitem [{\citenamefont {Llewellyn}\ \emph {et~al.}(2020)\citenamefont
  {Llewellyn}, \citenamefont {Bentley}, \citenamefont {Wadsworth},
  \citenamefont {Iwasaki}, \citenamefont {Dobaczewski}, \citenamefont
  {de~Angelis}, \citenamefont {Ash}, \citenamefont {Bazin}, \citenamefont
  {Bender}, \citenamefont {Cederwall}, \citenamefont {Crider}, \citenamefont
  {Doncel}, \citenamefont {Elder}, \citenamefont {Elman}, \citenamefont {Gade},
  \citenamefont {Grinder}, \citenamefont {Haylett}, \citenamefont {Jenkins},
  \citenamefont {Lee}, \citenamefont {Longfellow}, \citenamefont {Lunderberg},
  \citenamefont {Mijatovi\ifmmode~\acute{c}\else \'{c}\fi{}}, \citenamefont
  {Milne}, \citenamefont {Muir}, \citenamefont {Pastore}, \citenamefont
  {Rhodes},\ and\ \citenamefont {Weisshaar}}]{Llewellyn2020_PRL124-152501}%
  \BibitemOpen
  \bibfield  {author} {\bibinfo {author} {\bibfnamefont {R.~D.~O.}\
  \bibnamefont {Llewellyn}}, \bibinfo {author} {\bibfnamefont {M.~A.}\
  \bibnamefont {Bentley}}, \bibinfo {author} {\bibfnamefont {R.}~\bibnamefont
  {Wadsworth}}, \bibinfo {author} {\bibfnamefont {H.}~\bibnamefont {Iwasaki}},
  \bibinfo {author} {\bibfnamefont {J.}~\bibnamefont {Dobaczewski}}, \bibinfo
  {author} {\bibfnamefont {G.}~\bibnamefont {de~Angelis}}, \bibinfo {author}
  {\bibfnamefont {J.}~\bibnamefont {Ash}}, \bibinfo {author} {\bibfnamefont
  {D.}~\bibnamefont {Bazin}}, \bibinfo {author} {\bibfnamefont {P.~C.}\
  \bibnamefont {Bender}}, \bibinfo {author} {\bibfnamefont {B.}~\bibnamefont
  {Cederwall}}, \bibinfo {author} {\bibfnamefont {B.~P.}\ \bibnamefont
  {Crider}}, \bibinfo {author} {\bibfnamefont {M.}~\bibnamefont {Doncel}},
  \bibinfo {author} {\bibfnamefont {R.}~\bibnamefont {Elder}}, \bibinfo
  {author} {\bibfnamefont {B.}~\bibnamefont {Elman}}, \bibinfo {author}
  {\bibfnamefont {A.}~\bibnamefont {Gade}}, \bibinfo {author} {\bibfnamefont
  {M.}~\bibnamefont {Grinder}}, \bibinfo {author} {\bibfnamefont
  {T.}~\bibnamefont {Haylett}}, \bibinfo {author} {\bibfnamefont {D.~G.}\
  \bibnamefont {Jenkins}}, \bibinfo {author} {\bibfnamefont {I.~Y.}\
  \bibnamefont {Lee}}, \bibinfo {author} {\bibfnamefont {B.}~\bibnamefont
  {Longfellow}}, \bibinfo {author} {\bibfnamefont {E.}~\bibnamefont
  {Lunderberg}}, \bibinfo {author} {\bibfnamefont {T.}~\bibnamefont
  {Mijatovi\ifmmode~\acute{c}\else \'{c}\fi{}}}, \bibinfo {author}
  {\bibfnamefont {S.~A.}\ \bibnamefont {Milne}}, \bibinfo {author}
  {\bibfnamefont {D.}~\bibnamefont {Muir}}, \bibinfo {author} {\bibfnamefont
  {A.}~\bibnamefont {Pastore}}, \bibinfo {author} {\bibfnamefont
  {D.}~\bibnamefont {Rhodes}},\ and\ \bibinfo {author} {\bibfnamefont
  {D.}~\bibnamefont {Weisshaar}},\ }\href
  {https://doi.org/10.1103/PhysRevLett.124.152501} {\bibfield  {journal}
  {\bibinfo  {journal} {Phys. Rev. Lett.}\ }\textbf {\bibinfo {volume} {124}},\
  \bibinfo {pages} {152501} (\bibinfo {year} {2020})}\BibitemShut {NoStop}%
\bibitem [{\citenamefont {Hamaker}\ \emph {et~al.}(2021)\citenamefont
  {Hamaker}, \citenamefont {Leistenschneider}, \citenamefont {Jain},
  \citenamefont {Bollen}, \citenamefont {Giuliani}, \citenamefont {Lund},
  \citenamefont {Nazarewicz}, \citenamefont {Neufcourt}, \citenamefont
  {Nicoloff}, \citenamefont {Puentes}, \citenamefont {Ringle}, \citenamefont
  {Sumithrarachchi},\ and\ \citenamefont {Yandow}}]{Hamaker2021_NP17-1408}%
  \BibitemOpen
  \bibfield  {author} {\bibinfo {author} {\bibfnamefont {A.}~\bibnamefont
  {Hamaker}}, \bibinfo {author} {\bibfnamefont {E.}~\bibnamefont
  {Leistenschneider}}, \bibinfo {author} {\bibfnamefont {R.}~\bibnamefont
  {Jain}}, \bibinfo {author} {\bibfnamefont {G.}~\bibnamefont {Bollen}},
  \bibinfo {author} {\bibfnamefont {S.~A.}\ \bibnamefont {Giuliani}}, \bibinfo
  {author} {\bibfnamefont {K.}~\bibnamefont {Lund}}, \bibinfo {author}
  {\bibfnamefont {W.}~\bibnamefont {Nazarewicz}}, \bibinfo {author}
  {\bibfnamefont {L.}~\bibnamefont {Neufcourt}}, \bibinfo {author}
  {\bibfnamefont {C.~R.}\ \bibnamefont {Nicoloff}}, \bibinfo {author}
  {\bibfnamefont {D.}~\bibnamefont {Puentes}}, \bibinfo {author} {\bibfnamefont
  {R.}~\bibnamefont {Ringle}}, \bibinfo {author} {\bibfnamefont {C.~S.}\
  \bibnamefont {Sumithrarachchi}},\ and\ \bibinfo {author} {\bibfnamefont
  {I.~T.}\ \bibnamefont {Yandow}},\ }\href
  {https://doi.org/10.1038/s41567-021-01395-w} {\bibfield  {journal} {\bibinfo
  {journal} {Nature Phys.}\ }\textbf {\bibinfo {volume} {17}},\ \bibinfo
  {pages} {1408} (\bibinfo {year} {2021})}\BibitemShut {NoStop}%
\bibitem [{\citenamefont {Hiyama}\ \emph {et~al.}(1999)\citenamefont {Hiyama},
  \citenamefont {Kamimura}, \citenamefont {Miyazaki},\ and\ \citenamefont
  {Motoba}}]{Hiyama1999_PRC59-2351}%
  \BibitemOpen
  \bibfield  {author} {\bibinfo {author} {\bibfnamefont {E.}~\bibnamefont
  {Hiyama}}, \bibinfo {author} {\bibfnamefont {M.}~\bibnamefont {Kamimura}},
  \bibinfo {author} {\bibfnamefont {K.}~\bibnamefont {Miyazaki}},\ and\
  \bibinfo {author} {\bibfnamefont {T.}~\bibnamefont {Motoba}},\ }\href
  {https://doi.org/10.1103/PhysRevC.59.2351} {\bibfield  {journal} {\bibinfo
  {journal} {Phys. Rev. C}\ }\textbf {\bibinfo {volume} {59}},\ \bibinfo
  {pages} {2351} (\bibinfo {year} {1999})}\BibitemShut {NoStop}%
\bibitem [{\citenamefont {Tanida}\ \emph {et~al.}(2001)\citenamefont {Tanida},
  \citenamefont {Tamura}, \citenamefont {Abe}, \citenamefont {Akikawa},
  \citenamefont {Araki}, \citenamefont {Bhang} \emph
  {et~al.}}]{Tanida2001_PRL86-1982}%
  \BibitemOpen
  \bibfield  {author} {\bibinfo {author} {\bibfnamefont {K.}~\bibnamefont
  {Tanida}}, \bibinfo {author} {\bibfnamefont {H.}~\bibnamefont {Tamura}},
  \bibinfo {author} {\bibfnamefont {D.}~\bibnamefont {Abe}}, \bibinfo {author}
  {\bibfnamefont {H.}~\bibnamefont {Akikawa}}, \bibinfo {author} {\bibfnamefont
  {K.}~\bibnamefont {Araki}}, \bibinfo {author} {\bibfnamefont
  {H.}~\bibnamefont {Bhang}}, \emph {et~al.},\ }\href
  {https://doi.org/10.1103/PhysRevLett.86.1982} {\bibfield  {journal} {\bibinfo
   {journal} {Phys. Rev. Lett.}\ }\textbf {\bibinfo {volume} {86}},\ \bibinfo
  {pages} {1982} (\bibinfo {year} {2001})}\BibitemShut {NoStop}%
\bibitem [{\citenamefont {Isaka}\ \emph {et~al.}(2011)\citenamefont {Isaka},
  \citenamefont {Kimura}, \citenamefont {Dote},\ and\ \citenamefont
  {Ohnishi}}]{Isaka2011_PRC83-044323}%
  \BibitemOpen
  \bibfield  {author} {\bibinfo {author} {\bibfnamefont {M.}~\bibnamefont
  {Isaka}}, \bibinfo {author} {\bibfnamefont {M.}~\bibnamefont {Kimura}},
  \bibinfo {author} {\bibfnamefont {A.}~\bibnamefont {Dote}},\ and\ \bibinfo
  {author} {\bibfnamefont {A.}~\bibnamefont {Ohnishi}},\ }\href
  {https://doi.org/10.1103/PhysRevC.83.044323} {\bibfield  {journal} {\bibinfo
  {journal} {Phys. Rev. C}\ }\textbf {\bibinfo {volume} {83}},\ \bibinfo
  {pages} {044323} (\bibinfo {year} {2011})}\BibitemShut {NoStop}%
\bibitem [{\citenamefont {Isaka}\ and\ \citenamefont
  {Kimura}(2015)}]{Isaka2015_PRC92-044326}%
  \BibitemOpen
  \bibfield  {author} {\bibinfo {author} {\bibfnamefont {M.}~\bibnamefont
  {Isaka}}\ and\ \bibinfo {author} {\bibfnamefont {M.}~\bibnamefont {Kimura}},\
  }\href {https://doi.org/10.1103/PhysRevC.92.044326} {\bibfield  {journal}
  {\bibinfo  {journal} {Phys. Rev. C}\ }\textbf {\bibinfo {volume} {92}},\
  \bibinfo {pages} {044326} (\bibinfo {year} {2015})}\BibitemShut {NoStop}%
\bibitem [{\citenamefont {Tanimura}(2019)}]{Tanimura2019_PRC99-034324}%
  \BibitemOpen
  \bibfield  {author} {\bibinfo {author} {\bibfnamefont {Y.}~\bibnamefont
  {Tanimura}},\ }\href {https://doi.org/10.1103/PhysRevC.99.034324} {\bibfield
  {journal} {\bibinfo  {journal} {Phys. Rev. C}\ }\textbf {\bibinfo {volume}
  {99}},\ \bibinfo {pages} {034324} (\bibinfo {year} {2019})}\BibitemShut
  {NoStop}%
\bibitem [{\citenamefont {Hiyama}\ \emph {et~al.}(1996)\citenamefont {Hiyama},
  \citenamefont {Kamimura}, \citenamefont {Motoba}, \citenamefont {Yamada},\
  and\ \citenamefont {Yamamoto}}]{Hiyama1996_PRC53-2075}%
  \BibitemOpen
  \bibfield  {author} {\bibinfo {author} {\bibfnamefont {E.}~\bibnamefont
  {Hiyama}}, \bibinfo {author} {\bibfnamefont {M.}~\bibnamefont {Kamimura}},
  \bibinfo {author} {\bibfnamefont {T.}~\bibnamefont {Motoba}}, \bibinfo
  {author} {\bibfnamefont {T.}~\bibnamefont {Yamada}},\ and\ \bibinfo {author}
  {\bibfnamefont {Y.}~\bibnamefont {Yamamoto}},\ }\href
  {https://doi.org/10.1103/PhysRevC.53.2075} {\bibfield  {journal} {\bibinfo
  {journal} {Phys. Rev. C}\ }\textbf {\bibinfo {volume} {53}},\ \bibinfo
  {pages} {2075} (\bibinfo {year} {1996})}\BibitemShut {NoStop}%
\bibitem [{\citenamefont {L$\ddot{\text{u}}$}\ and\ \citenamefont
  {Meng}(2002)}]{Lu2002_CPL19-1775}%
  \BibitemOpen
  \bibfield  {author} {\bibinfo {author} {\bibfnamefont {H.-F.}\ \bibnamefont
  {L$\ddot{\text{u}}$}}\ and\ \bibinfo {author} {\bibfnamefont
  {J.}~\bibnamefont {Meng}},\ }\href
  {https://doi.org/10.1088/0256-307X/19/12/310} {\bibfield  {journal} {\bibinfo
   {journal} {Chin. Phys. Lett.}\ }\textbf {\bibinfo {volume} {19}},\ \bibinfo
  {eid} {1775} (\bibinfo {year} {2002})}\BibitemShut {NoStop}%
\bibitem [{\citenamefont {L$\ddot{\text{u}}$}\ \emph
  {et~al.}(2003)\citenamefont {L$\ddot{\text{u}}$}, \citenamefont {Meng},
  \citenamefont {Zhang},\ and\ \citenamefont {Zhou}}]{Lu2003_EPJA17-19}%
  \BibitemOpen
  \bibfield  {author} {\bibinfo {author} {\bibfnamefont {H.-F.}\ \bibnamefont
  {L$\ddot{\text{u}}$}}, \bibinfo {author} {\bibfnamefont {J.}~\bibnamefont
  {Meng}}, \bibinfo {author} {\bibfnamefont {S.-Q.}\ \bibnamefont {Zhang}},\
  and\ \bibinfo {author} {\bibfnamefont {S.-G.}\ \bibnamefont {Zhou}},\ }\href
  {https://doi.org/10.1140/epja/i2002-10136-3} {\bibfield  {journal} {\bibinfo
  {journal} {Eur. Phys. J. A}\ }\textbf {\bibinfo {volume} {17}},\ \bibinfo
  {pages} {19} (\bibinfo {year} {2003})}\BibitemShut {NoStop}%
\bibitem [{\citenamefont {Xue}\ \emph {et~al.}(2022)\citenamefont {Xue},
  \citenamefont {Chen}, \citenamefont {Zhou}, \citenamefont {Cheng},\ and\
  \citenamefont {Schulze}}]{Xue2022_PRC106-044306}%
  \BibitemOpen
  \bibfield  {author} {\bibinfo {author} {\bibfnamefont {H.-T.}\ \bibnamefont
  {Xue}}, \bibinfo {author} {\bibfnamefont {Q.~B.}\ \bibnamefont {Chen}},
  \bibinfo {author} {\bibfnamefont {X.-R.}\ \bibnamefont {Zhou}}, \bibinfo
  {author} {\bibfnamefont {Y.~Y.}\ \bibnamefont {Cheng}},\ and\ \bibinfo
  {author} {\bibfnamefont {H.-J.}\ \bibnamefont {Schulze}},\ }\href
  {https://doi.org/10.1103/PhysRevC.106.044306} {\bibfield  {journal} {\bibinfo
   {journal} {Phys. Rev. C}\ }\textbf {\bibinfo {volume} {106}},\ \bibinfo
  {pages} {044306} (\bibinfo {year} {2022})}\BibitemShut {NoStop}%
\bibitem [{\citenamefont {Zhang}\ \emph
  {et~al.}(2022{\natexlab{a}})\citenamefont {Zhang}, \citenamefont {Sagawa},\
  and\ \citenamefont {Hiyama}}]{Zhang2022_PTEP2022-023D01}%
  \BibitemOpen
  \bibfield  {author} {\bibinfo {author} {\bibfnamefont {Y.}~\bibnamefont
  {Zhang}}, \bibinfo {author} {\bibfnamefont {H.}~\bibnamefont {Sagawa}},\ and\
  \bibinfo {author} {\bibfnamefont {E.}~\bibnamefont {Hiyama}},\ }\href
  {https://doi.org/10.1093/ptep/ptac004} {\bibfield  {journal} {\bibinfo
  {journal} {Prog. Theor. Exp. Phys.}\ }\textbf {\bibinfo {volume} {2022}},\
  \bibinfo {pages} {023D01} (\bibinfo {year} {2022}{\natexlab{a}})}\BibitemShut
  {NoStop}%
\bibitem [{\citenamefont {Lu}\ \emph {et~al.}(2017)\citenamefont {Lu},
  \citenamefont {Liu}, \citenamefont {Ren}, \citenamefont {Zhang},\ and\
  \citenamefont {Sun}}]{Lu2017_JPG44-125104}%
  \BibitemOpen
  \bibfield  {author} {\bibinfo {author} {\bibfnamefont {W.-L.}\ \bibnamefont
  {Lu}}, \bibinfo {author} {\bibfnamefont {Z.-X.}\ \bibnamefont {Liu}},
  \bibinfo {author} {\bibfnamefont {S.-H.}\ \bibnamefont {Ren}}, \bibinfo
  {author} {\bibfnamefont {W.}~\bibnamefont {Zhang}},\ and\ \bibinfo {author}
  {\bibfnamefont {T.-T.}\ \bibnamefont {Sun}},\ }\href
  {https://doi.org/10.1088/1361-6471/aa8e2d} {\bibfield  {journal} {\bibinfo
  {journal} {J. Phys. G: Nucl. Part. Phys.}\ }\textbf {\bibinfo {volume}
  {44}},\ \bibinfo {pages} {125104} (\bibinfo {year} {2017})}\BibitemShut
  {NoStop}%
\bibitem [{\citenamefont {Samanta}\ \emph {et~al.}(2008)\citenamefont
  {Samanta}, \citenamefont {Chowdhury},\ and\ \citenamefont
  {Basu}}]{Samanta2008_JPG35-065101}%
  \BibitemOpen
  \bibfield  {author} {\bibinfo {author} {\bibfnamefont {C.}~\bibnamefont
  {Samanta}}, \bibinfo {author} {\bibfnamefont {P.}~\bibnamefont {Chowdhury}},\
  and\ \bibinfo {author} {\bibfnamefont {D.}~\bibnamefont {Basu}},\ }\href
  {https://doi.org/10.1088/0954-3899/35/6/065101} {\bibfield  {journal}
  {\bibinfo  {journal} {J. Phys. G: Nucl. Part. Phys.}\ }\textbf {\bibinfo
  {volume} {35}},\ \bibinfo {pages} {065101} (\bibinfo {year}
  {2008})}\BibitemShut {NoStop}%
\bibitem [{\citenamefont {Zhou}\ \emph {et~al.}(2008)\citenamefont {Zhou},
  \citenamefont {Polls}, \citenamefont {Schulze},\ and\ \citenamefont
  {Vida\~na}}]{Zhou2008_PRC78-054306}%
  \BibitemOpen
  \bibfield  {author} {\bibinfo {author} {\bibfnamefont {X.-R.}\ \bibnamefont
  {Zhou}}, \bibinfo {author} {\bibfnamefont {A.}~\bibnamefont {Polls}},
  \bibinfo {author} {\bibfnamefont {H.-J.}\ \bibnamefont {Schulze}},\ and\
  \bibinfo {author} {\bibfnamefont {I.}~\bibnamefont {Vida\~na}},\ }\href
  {https://doi.org/10.1103/PhysRevC.78.054306} {\bibfield  {journal} {\bibinfo
  {journal} {Phys. Rev. C}\ }\textbf {\bibinfo {volume} {78}},\ \bibinfo
  {pages} {054306} (\bibinfo {year} {2008})}\BibitemShut {NoStop}%
\bibitem [{\citenamefont {Wirth}\ and\ \citenamefont
  {Roth}(2018)}]{Wirth2018_PLB779-336}%
  \BibitemOpen
  \bibfield  {author} {\bibinfo {author} {\bibfnamefont {R.}~\bibnamefont
  {Wirth}}\ and\ \bibinfo {author} {\bibfnamefont {R.}~\bibnamefont {Roth}},\
  }\href {https://doi.org/10.1016/j.physletb.2018.02.021} {\bibfield  {journal}
  {\bibinfo  {journal} {Phys. Lett. B}\ }\textbf {\bibinfo {volume} {779}},\
  \bibinfo {pages} {336} (\bibinfo {year} {2018})}\BibitemShut {NoStop}%
\bibitem [{\citenamefont {Minato}\ \emph {et~al.}(2009)\citenamefont {Minato},
  \citenamefont {Chiba},\ and\ \citenamefont {Hagino}}]{Minato2009_NPA831-150}%
  \BibitemOpen
  \bibfield  {author} {\bibinfo {author} {\bibfnamefont {F.}~\bibnamefont
  {Minato}}, \bibinfo {author} {\bibfnamefont {S.}~\bibnamefont {Chiba}},\ and\
  \bibinfo {author} {\bibfnamefont {K.}~\bibnamefont {Hagino}},\ }\href
  {https://doi.org/https://doi.org/10.1016/j.nuclphysa.2009.09.063} {\bibfield
  {journal} {\bibinfo  {journal} {Nucl. Phys. A}\ }\textbf {\bibinfo {volume}
  {831}},\ \bibinfo {pages} {150 } (\bibinfo {year} {2009})}\BibitemShut
  {NoStop}%
\bibitem [{\citenamefont {Minato}\ and\ \citenamefont
  {Chiba}(2011)}]{Minato2011_NPA856-55}%
  \BibitemOpen
  \bibfield  {author} {\bibinfo {author} {\bibfnamefont {F.}~\bibnamefont
  {Minato}}\ and\ \bibinfo {author} {\bibfnamefont {S.}~\bibnamefont {Chiba}},\
  }\href {https://doi.org/https://doi.org/10.1016/j.nuclphysa.2011.02.127}
  {\bibfield  {journal} {\bibinfo  {journal} {Nucl. Phys. A}\ }\textbf
  {\bibinfo {volume} {856}},\ \bibinfo {pages} {55 } (\bibinfo {year}
  {2011})}\BibitemShut {NoStop}%
\bibitem [{\citenamefont {Zhou}\ \emph {et~al.}(2007)\citenamefont {Zhou},
  \citenamefont {Schulze}, \citenamefont {Sagawa}, \citenamefont {Wu},\ and\
  \citenamefont {Zhao}}]{Zhou2007_PRC76-034312}%
  \BibitemOpen
  \bibfield  {author} {\bibinfo {author} {\bibfnamefont {X.-R.}\ \bibnamefont
  {Zhou}}, \bibinfo {author} {\bibfnamefont {H.-J.}\ \bibnamefont {Schulze}},
  \bibinfo {author} {\bibfnamefont {H.}~\bibnamefont {Sagawa}}, \bibinfo
  {author} {\bibfnamefont {C.-X.}\ \bibnamefont {Wu}},\ and\ \bibinfo {author}
  {\bibfnamefont {E.-G.}\ \bibnamefont {Zhao}},\ }\href
  {https://doi.org/10.1103/PhysRevC.76.034312} {\bibfield  {journal} {\bibinfo
  {journal} {Phys. Rev. C}\ }\textbf {\bibinfo {volume} {76}},\ \bibinfo
  {pages} {034312} (\bibinfo {year} {2007})}\BibitemShut {NoStop}%
\bibitem [{\citenamefont {Win}\ and\ \citenamefont
  {Hagino}(2008)}]{Win2008_PRC78-054311}%
  \BibitemOpen
  \bibfield  {author} {\bibinfo {author} {\bibfnamefont {M.~T.}\ \bibnamefont
  {Win}}\ and\ \bibinfo {author} {\bibfnamefont {K.}~\bibnamefont {Hagino}},\
  }\href {https://doi.org/10.1103/PhysRevC.78.054311} {\bibfield  {journal}
  {\bibinfo  {journal} {Phys. Rev. C}\ }\textbf {\bibinfo {volume} {78}},\
  \bibinfo {pages} {054311} (\bibinfo {year} {2008})}\BibitemShut {NoStop}%
\bibitem [{\citenamefont {Lu}\ \emph {et~al.}(2014{\natexlab{a}})\citenamefont
  {Lu}, \citenamefont {Hiyama}, \citenamefont {Sagawa},\ and\ \citenamefont
  {Zhou}}]{Lu2014_PRC89-044307}%
  \BibitemOpen
  \bibfield  {author} {\bibinfo {author} {\bibfnamefont {B.-N.}\ \bibnamefont
  {Lu}}, \bibinfo {author} {\bibfnamefont {E.}~\bibnamefont {Hiyama}}, \bibinfo
  {author} {\bibfnamefont {H.}~\bibnamefont {Sagawa}},\ and\ \bibinfo {author}
  {\bibfnamefont {S.-G.}\ \bibnamefont {Zhou}},\ }\href
  {https://doi.org/10.1103/PhysRevC.89.044307} {\bibfield  {journal} {\bibinfo
  {journal} {Phys. Rev. C}\ }\textbf {\bibinfo {volume} {89}},\ \bibinfo
  {pages} {044307} (\bibinfo {year} {2014}{\natexlab{a}})}\BibitemShut
  {NoStop}%
\bibitem [{\citenamefont {Rong}\ \emph {et~al.}(2021)\citenamefont {Rong},
  \citenamefont {Tu},\ and\ \citenamefont {Zhou}}]{Rong2021_PRC104-054321}%
  \BibitemOpen
  \bibfield  {author} {\bibinfo {author} {\bibfnamefont {Y.-T.}\ \bibnamefont
  {Rong}}, \bibinfo {author} {\bibfnamefont {Z.-H.}\ \bibnamefont {Tu}},\ and\
  \bibinfo {author} {\bibfnamefont {S.-G.}\ \bibnamefont {Zhou}},\ }\href
  {https://doi.org/10.1103/physrevc.104.054321} {\bibfield  {journal} {\bibinfo
   {journal} {Phys. Rev. C}\ }\textbf {\bibinfo {volume} {104}},\ \bibinfo
  {pages} {054321} (\bibinfo {year} {2021})}\BibitemShut {NoStop}%
\bibitem [{\citenamefont {Li}\ \emph {et~al.}(2018)\citenamefont {Li},
  \citenamefont {Cui},\ and\ \citenamefont {Zhou}}]{Li2018_PRC97-034302}%
  \BibitemOpen
  \bibfield  {author} {\bibinfo {author} {\bibfnamefont {W.-Y.}\ \bibnamefont
  {Li}}, \bibinfo {author} {\bibfnamefont {J.-W.}\ \bibnamefont {Cui}},\ and\
  \bibinfo {author} {\bibfnamefont {X.-R.}\ \bibnamefont {Zhou}},\ }\href
  {https://doi.org/10.1103/PhysRevC.97.034302} {\bibfield  {journal} {\bibinfo
  {journal} {Phys. Rev. C}\ }\textbf {\bibinfo {volume} {97}},\ \bibinfo
  {pages} {034302} (\bibinfo {year} {2018})}\BibitemShut {NoStop}%
\bibitem [{\citenamefont {Chen}\ \emph {et~al.}(2021)\citenamefont {Chen},
  \citenamefont {Sun}, \citenamefont {Li},\ and\ \citenamefont
  {Sun}}]{Chen2021_SciChinaPMA64-282011}%
  \BibitemOpen
  \bibfield  {author} {\bibinfo {author} {\bibfnamefont {C.}~\bibnamefont
  {Chen}}, \bibinfo {author} {\bibfnamefont {Q.-K.}\ \bibnamefont {Sun}},
  \bibinfo {author} {\bibfnamefont {Y.-X.}\ \bibnamefont {Li}},\ and\ \bibinfo
  {author} {\bibfnamefont {T.-T.}\ \bibnamefont {Sun}},\ }\href
  {https://doi.org/10.1007/s11433-021-1721-1} {\bibfield  {journal} {\bibinfo
  {journal} {Sci. China-phys. Mech. Astron.}\ }\textbf {\bibinfo {volume}
  {64}},\ \bibinfo {pages} {282011} (\bibinfo {year} {2021})}\BibitemShut
  {NoStop}%
\bibitem [{\citenamefont {Chen}\ \emph {et~al.}(2022)\citenamefont {Chen},
  \citenamefont {Chen}, \citenamefont {Zhou}, \citenamefont {Cheng},
  \citenamefont {Cui},\ and\ \citenamefont {Schulze}}]{Chen2022_CPC46-064109}%
  \BibitemOpen
  \bibfield  {author} {\bibinfo {author} {\bibfnamefont {C.~F.}\ \bibnamefont
  {Chen}}, \bibinfo {author} {\bibfnamefont {Q.~B.}\ \bibnamefont {Chen}},
  \bibinfo {author} {\bibfnamefont {X.-R.}\ \bibnamefont {Zhou}}, \bibinfo
  {author} {\bibfnamefont {Y.~Y.}\ \bibnamefont {Cheng}}, \bibinfo {author}
  {\bibfnamefont {J.-W.}\ \bibnamefont {Cui}},\ and\ \bibinfo {author}
  {\bibfnamefont {H.-J.}\ \bibnamefont {Schulze}},\ }\href
  {https://doi.org/10.1088/1674-1137/ac5b58} {\bibfield  {journal} {\bibinfo
  {journal} {Chin. Phys. C}\ }\textbf {\bibinfo {volume} {46}},\ \bibinfo
  {pages} {064109} (\bibinfo {year} {2022})}\BibitemShut {NoStop}%
\bibitem [{\citenamefont {Lu}\ \emph {et~al.}(2011)\citenamefont {Lu},
  \citenamefont {Zhao},\ and\ \citenamefont {Zhou}}]{Lu2011_PRC84-014328}%
  \BibitemOpen
  \bibfield  {author} {\bibinfo {author} {\bibfnamefont {B.-N.}\ \bibnamefont
  {Lu}}, \bibinfo {author} {\bibfnamefont {E.-G.}\ \bibnamefont {Zhao}},\ and\
  \bibinfo {author} {\bibfnamefont {S.-G.}\ \bibnamefont {Zhou}},\ }\href
  {https://doi.org/10.1103/PhysRevC.84.014328} {\bibfield  {journal} {\bibinfo
  {journal} {Phys. Rev. C}\ }\textbf {\bibinfo {volume} {84}},\ \bibinfo
  {pages} {014328} (\bibinfo {year} {2011})}\BibitemShut {NoStop}%
\bibitem [{\citenamefont {Win}\ \emph {et~al.}(2011)\citenamefont {Win},
  \citenamefont {Hagino},\ and\ \citenamefont {Koike}}]{Win2011_PRC83-014301}%
  \BibitemOpen
  \bibfield  {author} {\bibinfo {author} {\bibfnamefont {M.~T.}\ \bibnamefont
  {Win}}, \bibinfo {author} {\bibfnamefont {K.}~\bibnamefont {Hagino}},\ and\
  \bibinfo {author} {\bibfnamefont {T.}~\bibnamefont {Koike}},\ }\href
  {https://doi.org/10.1103/PhysRevC.83.014301} {\bibfield  {journal} {\bibinfo
  {journal} {Phys. Rev. C}\ }\textbf {\bibinfo {volume} {83}},\ \bibinfo
  {pages} {014301} (\bibinfo {year} {2011})}\BibitemShut {NoStop}%
\bibitem [{\citenamefont {Yao}\ \emph {et~al.}(2011)\citenamefont {Yao},
  \citenamefont {Li}, \citenamefont {Hagino}, \citenamefont {Win},
  \citenamefont {Zhang},\ and\ \citenamefont {Meng}}]{Yao2011_NPA868-12}%
  \BibitemOpen
  \bibfield  {author} {\bibinfo {author} {\bibfnamefont {J.~M.}\ \bibnamefont
  {Yao}}, \bibinfo {author} {\bibfnamefont {Z.~P.}\ \bibnamefont {Li}},
  \bibinfo {author} {\bibfnamefont {K.}~\bibnamefont {Hagino}}, \bibinfo
  {author} {\bibfnamefont {M.}~\bibnamefont {Win}}, \bibinfo {author}
  {\bibfnamefont {Y.}~\bibnamefont {Zhang}},\ and\ \bibinfo {author}
  {\bibfnamefont {J.}~\bibnamefont {Meng}},\ }\href
  {https://doi.org/10.1016/j.nuclphysa.2011.08.006} {\bibfield  {journal}
  {\bibinfo  {journal} {Nucl. Phys. A}\ }\textbf {\bibinfo {volume}
  {868-869}},\ \bibinfo {pages} {12} (\bibinfo {year} {2011})}\BibitemShut
  {NoStop}%
\bibitem [{\citenamefont {Isaka}\ \emph {et~al.}(2013)\citenamefont {Isaka},
  \citenamefont {Kimura}, \citenamefont {Dot\'e},\ and\ \citenamefont
  {Ohnishi}}]{Isaka2013_PRC87-021304R}%
  \BibitemOpen
  \bibfield  {author} {\bibinfo {author} {\bibfnamefont {M.}~\bibnamefont
  {Isaka}}, \bibinfo {author} {\bibfnamefont {M.}~\bibnamefont {Kimura}},
  \bibinfo {author} {\bibfnamefont {A.}~\bibnamefont {Dot\'e}},\ and\ \bibinfo
  {author} {\bibfnamefont {A.}~\bibnamefont {Ohnishi}},\ }\href
  {https://doi.org/10.1103/PhysRevC.87.021304} {\bibfield  {journal} {\bibinfo
  {journal} {Phys. Rev. C}\ }\textbf {\bibinfo {volume} {87}},\ \bibinfo
  {pages} {021304(R)} (\bibinfo {year} {2013})}\BibitemShut {NoStop}%
\bibitem [{\citenamefont {Xue}\ \emph {et~al.}(2015)\citenamefont {Xue},
  \citenamefont {Yao}, \citenamefont {Hagino}, \citenamefont {Li},
  \citenamefont {Mei},\ and\ \citenamefont {Tanimura}}]{Xue2015_PRC91-024327}%
  \BibitemOpen
  \bibfield  {author} {\bibinfo {author} {\bibfnamefont {W.~X.}\ \bibnamefont
  {Xue}}, \bibinfo {author} {\bibfnamefont {J.~M.}\ \bibnamefont {Yao}},
  \bibinfo {author} {\bibfnamefont {K.}~\bibnamefont {Hagino}}, \bibinfo
  {author} {\bibfnamefont {Z.~P.}\ \bibnamefont {Li}}, \bibinfo {author}
  {\bibfnamefont {H.}~\bibnamefont {Mei}},\ and\ \bibinfo {author}
  {\bibfnamefont {Y.}~\bibnamefont {Tanimura}},\ }\href
  {https://doi.org/10.1103/PhysRevC.91.024327} {\bibfield  {journal} {\bibinfo
  {journal} {Phys. Rev. C}\ }\textbf {\bibinfo {volume} {91}},\ \bibinfo
  {pages} {024327} (\bibinfo {year} {2015})}\BibitemShut {NoStop}%
\bibitem [{\citenamefont {Cui}\ \emph {et~al.}(2015)\citenamefont {Cui},
  \citenamefont {Zhou},\ and\ \citenamefont {Schulze}}]{Cui2015_PRC91-054306}%
  \BibitemOpen
  \bibfield  {author} {\bibinfo {author} {\bibfnamefont {J.-W.}\ \bibnamefont
  {Cui}}, \bibinfo {author} {\bibfnamefont {X.-R.}\ \bibnamefont {Zhou}},\ and\
  \bibinfo {author} {\bibfnamefont {H.-J.}\ \bibnamefont {Schulze}},\ }\href
  {https://doi.org/10.1103/PhysRevC.91.054306} {\bibfield  {journal} {\bibinfo
  {journal} {Phys. Rev. C}\ }\textbf {\bibinfo {volume} {91}},\ \bibinfo
  {pages} {054306} (\bibinfo {year} {2015})}\BibitemShut {NoStop}%
\bibitem [{\citenamefont {Xia}\ \emph {et~al.}(2019)\citenamefont {Xia},
  \citenamefont {Wu}, \citenamefont {Mei},\ and\ \citenamefont
  {Yao}}]{Xia2019_SciChinaPMA62-042011}%
  \BibitemOpen
  \bibfield  {author} {\bibinfo {author} {\bibfnamefont {H.~J.}\ \bibnamefont
  {Xia}}, \bibinfo {author} {\bibfnamefont {X.~Y.}\ \bibnamefont {Wu}},
  \bibinfo {author} {\bibfnamefont {H.}~\bibnamefont {Mei}},\ and\ \bibinfo
  {author} {\bibfnamefont {J.~M.}\ \bibnamefont {Yao}},\ }\href
  {https://doi.org/10.1007/s11433-018-9308-0} {\bibfield  {journal} {\bibinfo
  {journal} {Sci. China-phys. Mech. Astron.}\ }\textbf {\bibinfo {volume}
  {62}},\ \bibinfo {pages} {42011} (\bibinfo {year} {2019})}\BibitemShut
  {NoStop}%
\bibitem [{\citenamefont {Lu}\ \emph {et~al.}(2012)\citenamefont {Lu},
  \citenamefont {Zhao},\ and\ \citenamefont {Zhou}}]{Lu2012_PRC85-011301R}%
  \BibitemOpen
  \bibfield  {author} {\bibinfo {author} {\bibfnamefont {B.-N.}\ \bibnamefont
  {Lu}}, \bibinfo {author} {\bibfnamefont {E.-G.}\ \bibnamefont {Zhao}},\ and\
  \bibinfo {author} {\bibfnamefont {S.-G.}\ \bibnamefont {Zhou}},\ }\href
  {https://doi.org/10.1103/PhysRevC.85.011301} {\bibfield  {journal} {\bibinfo
  {journal} {Phys. Rev. C}\ }\textbf {\bibinfo {volume} {85}},\ \bibinfo
  {pages} {011301(R)} (\bibinfo {year} {2012})}\BibitemShut {NoStop}%
\bibitem [{\citenamefont {Lu}\ \emph {et~al.}(2014{\natexlab{b}})\citenamefont
  {Lu}, \citenamefont {Zhao}, \citenamefont {Zhao},\ and\ \citenamefont
  {Zhou}}]{Lu2014_PRC89-014323}%
  \BibitemOpen
  \bibfield  {author} {\bibinfo {author} {\bibfnamefont {B.-N.}\ \bibnamefont
  {Lu}}, \bibinfo {author} {\bibfnamefont {J.}~\bibnamefont {Zhao}}, \bibinfo
  {author} {\bibfnamefont {E.-G.}\ \bibnamefont {Zhao}},\ and\ \bibinfo
  {author} {\bibfnamefont {S.-G.}\ \bibnamefont {Zhou}},\ }\href
  {https://doi.org/10.1103/PhysRevC.89.014323} {\bibfield  {journal} {\bibinfo
  {journal} {Phys. Rev. C}\ }\textbf {\bibinfo {volume} {89}},\ \bibinfo
  {pages} {014323} (\bibinfo {year} {2014}{\natexlab{b}})}\BibitemShut
  {NoStop}%
\bibitem [{\citenamefont {Zhou}(2016)}]{Zhou2016_PS91-063008}%
  \BibitemOpen
  \bibfield  {author} {\bibinfo {author} {\bibfnamefont {S.-G.}\ \bibnamefont
  {Zhou}},\ }\href {https://doi.org/10.1088/0031-8949/91/6/063008} {\bibfield
  {journal} {\bibinfo  {journal} {Phys. Scr.}\ }\textbf {\bibinfo {volume}
  {91}},\ \bibinfo {pages} {063008} (\bibinfo {year} {2016})}\BibitemShut
  {NoStop}%
\bibitem [{\citenamefont {Zhao}\ \emph {et~al.}(2017)\citenamefont {Zhao},
  \citenamefont {Lu}, \citenamefont {Zhao},\ and\ \citenamefont
  {Zhou}}]{Zhao2017_PRC95-014320}%
  \BibitemOpen
  \bibfield  {author} {\bibinfo {author} {\bibfnamefont {J.}~\bibnamefont
  {Zhao}}, \bibinfo {author} {\bibfnamefont {B.-N.}\ \bibnamefont {Lu}},
  \bibinfo {author} {\bibfnamefont {E.-G.}\ \bibnamefont {Zhao}},\ and\
  \bibinfo {author} {\bibfnamefont {S.-G.}\ \bibnamefont {Zhou}},\ }\href
  {https://doi.org/10.1103/PhysRevC.95.014320} {\bibfield  {journal} {\bibinfo
  {journal} {Phys. Rev. C}\ }\textbf {\bibinfo {volume} {95}},\ \bibinfo
  {pages} {014320} (\bibinfo {year} {2017})}\BibitemShut {NoStop}%
\bibitem [{\citenamefont {Zhao}\ \emph {et~al.}(2012)\citenamefont {Zhao},
  \citenamefont {Lu}, \citenamefont {Zhao},\ and\ \citenamefont
  {Zhou}}]{Zhao2012_PRC86-057304}%
  \BibitemOpen
  \bibfield  {author} {\bibinfo {author} {\bibfnamefont {J.}~\bibnamefont
  {Zhao}}, \bibinfo {author} {\bibfnamefont {B.-N.}\ \bibnamefont {Lu}},
  \bibinfo {author} {\bibfnamefont {E.-G.}\ \bibnamefont {Zhao}},\ and\
  \bibinfo {author} {\bibfnamefont {S.-G.}\ \bibnamefont {Zhou}},\ }\href
  {https://doi.org/10.1103/PhysRevC.86.057304} {\bibfield  {journal} {\bibinfo
  {journal} {Phys. Rev. C}\ }\textbf {\bibinfo {volume} {86}},\ \bibinfo
  {pages} {057304} (\bibinfo {year} {2012})}\BibitemShut {NoStop}%
\bibitem [{\citenamefont {Rong}\ \emph {et~al.}(2020)\citenamefont {Rong},
  \citenamefont {Zhao},\ and\ \citenamefont {Zhou}}]{Rong2020_PLB807-135533}%
  \BibitemOpen
  \bibfield  {author} {\bibinfo {author} {\bibfnamefont {Y.-T.}\ \bibnamefont
  {Rong}}, \bibinfo {author} {\bibfnamefont {P.}~\bibnamefont {Zhao}},\ and\
  \bibinfo {author} {\bibfnamefont {S.-G.}\ \bibnamefont {Zhou}},\ }\href
  {https://doi.org/10.1016/j.physletb.2020.135533} {\bibfield  {journal}
  {\bibinfo  {journal} {Phys. Lett. B}\ }\textbf {\bibinfo {volume} {807}},\
  \bibinfo {pages} {135533} (\bibinfo {year} {2020})}\BibitemShut {NoStop}%
\bibitem [{\citenamefont {Reinhard}(1989)}]{Reinhard1989_RPP52-439}%
  \BibitemOpen
  \bibfield  {author} {\bibinfo {author} {\bibfnamefont {P.~G.}\ \bibnamefont
  {Reinhard}},\ }\href {https://doi.org/10.1088/0034-4885/52/4/002} {\bibfield
  {journal} {\bibinfo  {journal} {Rep. Prog. Phys.}\ }\textbf {\bibinfo
  {volume} {52}},\ \bibinfo {pages} {439} (\bibinfo {year} {1989})}\BibitemShut
  {NoStop}%
\bibitem [{\citenamefont {Ring}(1996)}]{Ring1996_PPNP37-193}%
  \BibitemOpen
  \bibfield  {author} {\bibinfo {author} {\bibfnamefont {P.}~\bibnamefont
  {Ring}},\ }\href {https://doi.org/10.1016/0146-6410(96)00054-3} {\bibfield
  {journal} {\bibinfo  {journal} {Prog. Part. Nucl. Phys.}\ }\textbf {\bibinfo
  {volume} {37}},\ \bibinfo {pages} {193} (\bibinfo {year} {1996})}\BibitemShut
  {NoStop}%
\bibitem [{\citenamefont {Bender}\ \emph {et~al.}(2003)\citenamefont {Bender},
  \citenamefont {Heenen},\ and\ \citenamefont
  {Reinhard}}]{Bender2003_RMP75-121}%
  \BibitemOpen
  \bibfield  {author} {\bibinfo {author} {\bibfnamefont {M.}~\bibnamefont
  {Bender}}, \bibinfo {author} {\bibfnamefont {P.-H.}\ \bibnamefont {Heenen}},\
  and\ \bibinfo {author} {\bibfnamefont {P.-G.}\ \bibnamefont {Reinhard}},\
  }\href {https://doi.org/10.1103/RevModPhys.75.121} {\bibfield  {journal}
  {\bibinfo  {journal} {Rev. Mod. Phys.}\ }\textbf {\bibinfo {volume} {75}},\
  \bibinfo {pages} {121} (\bibinfo {year} {2003})}\BibitemShut {NoStop}%
\bibitem [{\citenamefont {Vretenar}\ \emph {et~al.}(2005)\citenamefont
  {Vretenar}, \citenamefont {Afanasjev}, \citenamefont {Lalazissis},\ and\
  \citenamefont {Ring}}]{Vretenar2005_PR409-101}%
  \BibitemOpen
  \bibfield  {author} {\bibinfo {author} {\bibfnamefont {D.}~\bibnamefont
  {Vretenar}}, \bibinfo {author} {\bibfnamefont {A.}~\bibnamefont {Afanasjev}},
  \bibinfo {author} {\bibfnamefont {G.}~\bibnamefont {Lalazissis}},\ and\
  \bibinfo {author} {\bibfnamefont {P.}~\bibnamefont {Ring}},\ }\href
  {https://doi.org/10.1016/j.physrep.2004.10.001} {\bibfield  {journal}
  {\bibinfo  {journal} {Phys. Rep.}\ }\textbf {\bibinfo {volume} {409}},\
  \bibinfo {pages} {101} (\bibinfo {year} {2005})}\BibitemShut {NoStop}%
\bibitem [{\citenamefont {Meng}\ \emph {et~al.}(2006)\citenamefont {Meng},
  \citenamefont {Toki}, \citenamefont {Zhou}, \citenamefont {Zhang},
  \citenamefont {Long},\ and\ \citenamefont {Geng}}]{Meng2006_PPNP57-470}%
  \BibitemOpen
  \bibfield  {author} {\bibinfo {author} {\bibfnamefont {J.}~\bibnamefont
  {Meng}}, \bibinfo {author} {\bibfnamefont {H.}~\bibnamefont {Toki}}, \bibinfo
  {author} {\bibfnamefont {S.~G.}\ \bibnamefont {Zhou}}, \bibinfo {author}
  {\bibfnamefont {S.~Q.}\ \bibnamefont {Zhang}}, \bibinfo {author}
  {\bibfnamefont {W.~H.}\ \bibnamefont {Long}},\ and\ \bibinfo {author}
  {\bibfnamefont {L.~S.}\ \bibnamefont {Geng}},\ }\href
  {https://doi.org/10.1016/j.ppnp.2005.06.001} {\bibfield  {journal} {\bibinfo
  {journal} {Prog. Part. Nucl. Phys.}\ }\textbf {\bibinfo {volume} {57}},\
  \bibinfo {pages} {470} (\bibinfo {year} {2006})}\BibitemShut {NoStop}%
\bibitem [{\citenamefont {Nik$\check{\text{s}}$i$\acute{\text{c}}$}\ \emph
  {et~al.}(2011)\citenamefont {Nik$\check{\text{s}}$i$\acute{\text{c}}$},
  \citenamefont {Vretenar},\ and\ \citenamefont
  {Ring}}]{Niksic2011_PPNP66-519}%
  \BibitemOpen
  \bibfield  {author} {\bibinfo {author} {\bibfnamefont {T.}~\bibnamefont
  {Nik$\check{\text{s}}$i$\acute{\text{c}}$}}, \bibinfo {author} {\bibfnamefont
  {D.}~\bibnamefont {Vretenar}},\ and\ \bibinfo {author} {\bibfnamefont
  {P.}~\bibnamefont {Ring}},\ }\href
  {https://doi.org/10.1016/j.ppnp.2011.01.055} {\bibfield  {journal} {\bibinfo
  {journal} {Prog. Part. Nucl. Phys.}\ }\textbf {\bibinfo {volume} {66}},\
  \bibinfo {pages} {519} (\bibinfo {year} {2011})}\BibitemShut {NoStop}%
\bibitem [{\citenamefont {Liang}\ \emph {et~al.}(2015)\citenamefont {Liang},
  \citenamefont {Meng},\ and\ \citenamefont {Zhou}}]{Liang2015_PR570-1}%
  \BibitemOpen
  \bibfield  {author} {\bibinfo {author} {\bibfnamefont {H.}~\bibnamefont
  {Liang}}, \bibinfo {author} {\bibfnamefont {J.}~\bibnamefont {Meng}},\ and\
  \bibinfo {author} {\bibfnamefont {S.-G.}\ \bibnamefont {Zhou}},\ }\href
  {https://doi.org/10.1016/j.physrep.2014.12.005} {\bibfield  {journal}
  {\bibinfo  {journal} {Phys. Rep.}\ }\textbf {\bibinfo {volume} {570}},\
  \bibinfo {pages} {1} (\bibinfo {year} {2015})}\BibitemShut {NoStop}%
\bibitem [{\citenamefont {Meng}\ and\ \citenamefont
  {Zhou}(2015)}]{Meng2015_JPG42-093101}%
  \BibitemOpen
  \bibfield  {author} {\bibinfo {author} {\bibfnamefont {J.}~\bibnamefont
  {Meng}}\ and\ \bibinfo {author} {\bibfnamefont {S.-G.}\ \bibnamefont
  {Zhou}},\ }\href {https://doi.org/10.1088/0954-3899/42/9/093101} {\bibfield
  {journal} {\bibinfo  {journal} {J. Phys. G: Nucl. Part. Phys.}\ }\textbf
  {\bibinfo {volume} {42}},\ \bibinfo {pages} {093101} (\bibinfo {year}
  {2015})}\BibitemShut {NoStop}%
\bibitem [{\citenamefont {Meng}(2016)}]{Meng2016_RDFNS}%
  \BibitemOpen
  \bibinfo {editor} {\bibfnamefont {J.}~\bibnamefont {Meng}},\ ed.,\ \href
  {https://doi.org/10.1142/9872} {\emph {\bibinfo {title} {Relativistic
  {D}ensity {F}unctional for {N}uclear {S}tructure}}},\ \bibinfo {edition}
  {{V}ol. 10 of {I}nternational {R}eview of {N}uclear {P}hysics}\ ed.\
  (\bibinfo  {publisher} {World Scientific Pub Co Pte Lt},\ \bibinfo {year}
  {2016})\BibitemShut {NoStop}%
\bibitem [{\citenamefont {Schaffner}\ \emph {et~al.}(1994)\citenamefont
  {Schaffner}, \citenamefont {Dover}, \citenamefont {Gal}, \citenamefont
  {Greiner}, \citenamefont {Millener},\ and\ \citenamefont
  {Stocker}}]{Schaffner1994_AP235-35}%
  \BibitemOpen
  \bibfield  {author} {\bibinfo {author} {\bibfnamefont {J.}~\bibnamefont
  {Schaffner}}, \bibinfo {author} {\bibfnamefont {C.}~\bibnamefont {Dover}},
  \bibinfo {author} {\bibfnamefont {A.}~\bibnamefont {Gal}}, \bibinfo {author}
  {\bibfnamefont {C.}~\bibnamefont {Greiner}}, \bibinfo {author} {\bibfnamefont
  {D.}~\bibnamefont {Millener}},\ and\ \bibinfo {author} {\bibfnamefont
  {H.}~\bibnamefont {Stocker}},\ }\href
  {https://doi.org/10.1006/aphy.1994.1090} {\bibfield  {journal} {\bibinfo
  {journal} {Ann. Phys.}\ }\textbf {\bibinfo {volume} {235}},\ \bibinfo {pages}
  {35} (\bibinfo {year} {1994})}\BibitemShut {NoStop}%
\bibitem [{\citenamefont {Shen}\ \emph {et~al.}(2006)\citenamefont {Shen},
  \citenamefont {Yang},\ and\ \citenamefont {Toki}}]{Shen2006_PTP115-325}%
  \BibitemOpen
  \bibfield  {author} {\bibinfo {author} {\bibfnamefont {H.}~\bibnamefont
  {Shen}}, \bibinfo {author} {\bibfnamefont {F.}~\bibnamefont {Yang}},\ and\
  \bibinfo {author} {\bibfnamefont {H.}~\bibnamefont {Toki}},\ }\href
  {https://doi.org/10.1143/PTP.115.325} {\bibfield  {journal} {\bibinfo
  {journal} {Prog. Theor. Phys.}\ }\textbf {\bibinfo {volume} {115}},\ \bibinfo
  {pages} {325} (\bibinfo {year} {2006})}\BibitemShut {NoStop}%
\bibitem [{\citenamefont {Schaffner}\ and\ \citenamefont
  {Mishustin}(1996)}]{Schaffner1996_PRC53-1416}%
  \BibitemOpen
  \bibfield  {author} {\bibinfo {author} {\bibfnamefont {J.}~\bibnamefont
  {Schaffner}}\ and\ \bibinfo {author} {\bibfnamefont {I.~N.}\ \bibnamefont
  {Mishustin}},\ }\href {https://doi.org/10.1103/PhysRevC.53.1416} {\bibfield
  {journal} {\bibinfo  {journal} {Phys. Rev. C}\ }\textbf {\bibinfo {volume}
  {53}},\ \bibinfo {pages} {1416} (\bibinfo {year} {1996})}\BibitemShut
  {NoStop}%
\bibitem [{\citenamefont {Weissenborn}\ \emph {et~al.}(2013)\citenamefont
  {Weissenborn}, \citenamefont {Chatterjee},\ and\ \citenamefont
  {Schaffner-Bielich}}]{Weissenborn2013_NPA914-421}%
  \BibitemOpen
  \bibfield  {author} {\bibinfo {author} {\bibfnamefont {S.}~\bibnamefont
  {Weissenborn}}, \bibinfo {author} {\bibfnamefont {D.}~\bibnamefont
  {Chatterjee}},\ and\ \bibinfo {author} {\bibfnamefont {J.}~\bibnamefont
  {Schaffner-Bielich}},\ }\href
  {https://doi.org/10.1016/j.nuclphysa.2013.04.003} {\bibfield  {journal}
  {\bibinfo  {journal} {Nucl. Phys. A}\ }\textbf {\bibinfo {volume} {914}},\
  \bibinfo {pages} {421} (\bibinfo {year} {2013})}\BibitemShut {NoStop}%
\bibitem [{\citenamefont {Tu}\ and\ \citenamefont
  {Zhou}(2022)}]{Tu2022_ApJ925-16}%
  \BibitemOpen
  \bibfield  {author} {\bibinfo {author} {\bibfnamefont {Z.-H.}\ \bibnamefont
  {Tu}}\ and\ \bibinfo {author} {\bibfnamefont {S.-G.}\ \bibnamefont {Zhou}},\
  }\href {https://doi.org/10.3847/1538-4357/ac3996} {\bibfield  {journal}
  {\bibinfo  {journal} {Astrophys. J.}\ }\textbf {\bibinfo {volume} {925}},\
  \bibinfo {pages} {16} (\bibinfo {year} {2022})}\BibitemShut {NoStop}%
\bibitem [{\citenamefont {Boguta}\ and\ \citenamefont
  {Bodmer}(1977)}]{Boguta1977_NPA292-413}%
  \BibitemOpen
  \bibfield  {author} {\bibinfo {author} {\bibfnamefont {J.}~\bibnamefont
  {Boguta}}\ and\ \bibinfo {author} {\bibfnamefont {A.}~\bibnamefont
  {Bodmer}},\ }\href {https://doi.org/10.1016/0375-9474(77)90626-1} {\bibfield
  {journal} {\bibinfo  {journal} {Nucl. Phys. A}\ }\textbf {\bibinfo {volume}
  {292}},\ \bibinfo {pages} {413} (\bibinfo {year} {1977})}\BibitemShut
  {NoStop}%
\bibitem [{\citenamefont {Sugahara}\ and\ \citenamefont
  {Toki}(1994)}]{Sugahara1994_NPA579-557}%
  \BibitemOpen
  \bibfield  {author} {\bibinfo {author} {\bibfnamefont {Y.}~\bibnamefont
  {Sugahara}}\ and\ \bibinfo {author} {\bibfnamefont {H.}~\bibnamefont
  {Toki}},\ }\href {https://doi.org/10.1016/0375-9474(94)90923-7} {\bibfield
  {journal} {\bibinfo  {journal} {Nucl. Phys. A}\ }\textbf {\bibinfo {volume}
  {579}},\ \bibinfo {pages} {557} (\bibinfo {year} {1994})}\BibitemShut
  {NoStop}%
\bibitem [{\citenamefont {Long}\ \emph {et~al.}(2004)\citenamefont {Long},
  \citenamefont {Meng}, \citenamefont {{N. Van Giai}},\ and\ \citenamefont
  {Zhou}}]{Long2004_PRC69-034319}%
  \BibitemOpen
  \bibfield  {author} {\bibinfo {author} {\bibfnamefont {W.}~\bibnamefont
  {Long}}, \bibinfo {author} {\bibfnamefont {J.}~\bibnamefont {Meng}}, \bibinfo
  {author} {\bibnamefont {{N. Van Giai}}},\ and\ \bibinfo {author}
  {\bibfnamefont {S.~G.}\ \bibnamefont {Zhou}},\ }\href
  {https://doi.org/10.1103/PhysRevC.69.034319} {\bibfield  {journal} {\bibinfo
  {journal} {Phys. Rev. C}\ }\textbf {\bibinfo {volume} {69}},\ \bibinfo
  {pages} {034319} (\bibinfo {year} {2004})}\BibitemShut {NoStop}%
\bibitem [{\citenamefont {Typel}\ and\ \citenamefont
  {Wolter}(1999)}]{Typel1999_NPA656-331}%
  \BibitemOpen
  \bibfield  {author} {\bibinfo {author} {\bibfnamefont {S.}~\bibnamefont
  {Typel}}\ and\ \bibinfo {author} {\bibfnamefont {H.~H.}\ \bibnamefont
  {Wolter}},\ }\href {https://doi.org/10.1016/S0375-9474(99)00310-3} {\bibfield
   {journal} {\bibinfo  {journal} {Nucl. Phys. A}\ }\textbf {\bibinfo {volume}
  {656}},\ \bibinfo {pages} {331} (\bibinfo {year} {1999})}\BibitemShut
  {NoStop}%
\bibitem [{\citenamefont {Nik\ifmmode \check{s}\else
  \v{s}\fi{}i\ifmmode~\acute{c}\else \'{c}\fi{}}\ \emph
  {et~al.}(2002)\citenamefont {Nik\ifmmode \check{s}\else
  \v{s}\fi{}i\ifmmode~\acute{c}\else \'{c}\fi{}}, \citenamefont {Vretenar},
  \citenamefont {Finelli},\ and\ \citenamefont
  {Ring}}]{Niksic2002_PRC66-024306}%
  \BibitemOpen
  \bibfield  {author} {\bibinfo {author} {\bibfnamefont {T.}~\bibnamefont
  {Nik\ifmmode \check{s}\else \v{s}\fi{}i\ifmmode~\acute{c}\else \'{c}\fi{}}},
  \bibinfo {author} {\bibfnamefont {D.}~\bibnamefont {Vretenar}}, \bibinfo
  {author} {\bibfnamefont {P.}~\bibnamefont {Finelli}},\ and\ \bibinfo {author}
  {\bibfnamefont {P.}~\bibnamefont {Ring}},\ }\href
  {https://doi.org/10.1103/PhysRevC.66.024306} {\bibfield  {journal} {\bibinfo
  {journal} {Phys. Rev. C}\ }\textbf {\bibinfo {volume} {66}},\ \bibinfo
  {pages} {024306} (\bibinfo {year} {2002})}\BibitemShut {NoStop}%
\bibitem [{\citenamefont {Lalazissis}\ \emph {et~al.}(2005)\citenamefont
  {Lalazissis}, \citenamefont {Nik\ifmmode \check{s}\else
  \v{s}\fi{}i\ifmmode~\acute{c}\else \'{c}\fi{}}, \citenamefont {Vretenar},\
  and\ \citenamefont {Ring}}]{Lalazissis2005_PRC71-024312}%
  \BibitemOpen
  \bibfield  {author} {\bibinfo {author} {\bibfnamefont {G.~A.}\ \bibnamefont
  {Lalazissis}}, \bibinfo {author} {\bibfnamefont {T.}~\bibnamefont
  {Nik\ifmmode \check{s}\else \v{s}\fi{}i\ifmmode~\acute{c}\else \'{c}\fi{}}},
  \bibinfo {author} {\bibfnamefont {D.}~\bibnamefont {Vretenar}},\ and\
  \bibinfo {author} {\bibfnamefont {P.}~\bibnamefont {Ring}},\ }\href
  {https://doi.org/10.1103/PhysRevC.71.024312} {\bibfield  {journal} {\bibinfo
  {journal} {Phys. Rev. C}\ }\textbf {\bibinfo {volume} {71}},\ \bibinfo
  {pages} {024312} (\bibinfo {year} {2005})}\BibitemShut {NoStop}%
\bibitem [{\citenamefont {Wang}\ \emph {et~al.}(2013)\citenamefont {Wang},
  \citenamefont {Sang}, \citenamefont {Wang},\ and\ \citenamefont
  {L$\mathrm{\ddot{u}}$}}]{Wang2013_CTP60-479}%
  \BibitemOpen
  \bibfield  {author} {\bibinfo {author} {\bibfnamefont {X.-S.}\ \bibnamefont
  {Wang}}, \bibinfo {author} {\bibfnamefont {H.-Y.}\ \bibnamefont {Sang}},
  \bibinfo {author} {\bibfnamefont {J.-H.}\ \bibnamefont {Wang}},\ and\
  \bibinfo {author} {\bibfnamefont {H.-F.}\ \bibnamefont
  {L$\mathrm{\ddot{u}}$}},\ }\href {https://doi.org/10.1088/0253-6102/60/4/16}
  {\bibfield  {journal} {\bibinfo  {journal} {Commun. Theor. Phys.}\ }\textbf
  {\bibinfo {volume} {60}},\ \bibinfo {pages} {479} (\bibinfo {year}
  {2013})}\BibitemShut {NoStop}%
\bibitem [{\citenamefont {Sun}\ \emph {et~al.}(2018)\citenamefont {Sun},
  \citenamefont {Xia}, \citenamefont {Zhang},\ and\ \citenamefont
  {Smith}}]{Sun2018_CPC42-25101}%
  \BibitemOpen
  \bibfield  {author} {\bibinfo {author} {\bibfnamefont {T.-T.}\ \bibnamefont
  {Sun}}, \bibinfo {author} {\bibfnamefont {C.-J.}\ \bibnamefont {Xia}},
  \bibinfo {author} {\bibfnamefont {S.-S.}\ \bibnamefont {Zhang}},\ and\
  \bibinfo {author} {\bibfnamefont {M.~S.}\ \bibnamefont {Smith}},\ }\href
  {https://doi.org/10.1088/1674-1137/42/2/025101} {\bibfield  {journal}
  {\bibinfo  {journal} {Chin. Phys. C}\ }\textbf {\bibinfo {volume} {42}},\
  \bibinfo {eid} {25101} (\bibinfo {year} {2018})}\BibitemShut {NoStop}%
\bibitem [{\citenamefont {{van Dalen}}\ \emph {et~al.}(2014)\citenamefont {{van
  Dalen}}, \citenamefont {Colucci},\ and\ \citenamefont
  {Sedrakian}}]{van-Dalen2014_PLB734-383}%
  \BibitemOpen
  \bibfield  {author} {\bibinfo {author} {\bibfnamefont {E.}~\bibnamefont {{van
  Dalen}}}, \bibinfo {author} {\bibfnamefont {G.}~\bibnamefont {Colucci}},\
  and\ \bibinfo {author} {\bibfnamefont {A.}~\bibnamefont {Sedrakian}},\ }\href
  {https://doi.org/10.1016/j.physletb.2014.06.002} {\bibfield  {journal}
  {\bibinfo  {journal} {Phys. Lett. B}\ }\textbf {\bibinfo {volume} {734}},\
  \bibinfo {pages} {383} (\bibinfo {year} {2014})}\BibitemShut {NoStop}%
\bibitem [{\citenamefont {Lenske}\ and\ \citenamefont
  {Fuchs}(1995)}]{Lenske1995_PLB345-355}%
  \BibitemOpen
  \bibfield  {author} {\bibinfo {author} {\bibfnamefont {H.}~\bibnamefont
  {Lenske}}\ and\ \bibinfo {author} {\bibfnamefont {C.}~\bibnamefont {Fuchs}},\
  }\href {https://doi.org/10.1016/0370-2693(94)01664-X} {\bibfield  {journal}
  {\bibinfo  {journal} {Phys. Lett. B}\ }\textbf {\bibinfo {volume} {345}},\
  \bibinfo {pages} {355} (\bibinfo {year} {1995})}\BibitemShut {NoStop}%
\bibitem [{\citenamefont {Tian}\ and\ \citenamefont
  {Ma}(2006)}]{Tian2006_CPL23-3226}%
  \BibitemOpen
  \bibfield  {author} {\bibinfo {author} {\bibfnamefont {Y.}~\bibnamefont
  {Tian}}\ and\ \bibinfo {author} {\bibfnamefont {Z.-Y.}\ \bibnamefont {Ma}},\
  }\href {https://cpl.iphy.ac.cn/Y2006/V23/I12/3226} {\bibfield  {journal}
  {\bibinfo  {journal} {Chin. Phys. Lett.}\ }\textbf {\bibinfo {volume} {23}},\
  \bibinfo {eid} {3226-3229} (\bibinfo {year} {2006})}\BibitemShut {NoStop}%
\bibitem [{\citenamefont {Tian}\ \emph {et~al.}(2009)\citenamefont {Tian},
  \citenamefont {Ma},\ and\ \citenamefont {Ring}}]{Tian2009_PLB676-44}%
  \BibitemOpen
  \bibfield  {author} {\bibinfo {author} {\bibfnamefont {Y.}~\bibnamefont
  {Tian}}, \bibinfo {author} {\bibfnamefont {Z.~Y.}\ \bibnamefont {Ma}},\ and\
  \bibinfo {author} {\bibfnamefont {P.}~\bibnamefont {Ring}},\ }\href
  {https://doi.org/10.1016/j.physletb.2009.04.067} {\bibfield  {journal}
  {\bibinfo  {journal} {Phys. Lett. B}\ }\textbf {\bibinfo {volume} {676}},\
  \bibinfo {pages} {44} (\bibinfo {year} {2009})}\BibitemShut {NoStop}%
\bibitem [{\citenamefont {Perez-Martin}\ and\ \citenamefont
  {Robledo}(2008)}]{Perez-Martin2008_PRC78-014304}%
  \BibitemOpen
  \bibfield  {author} {\bibinfo {author} {\bibfnamefont {S.}~\bibnamefont
  {Perez-Martin}}\ and\ \bibinfo {author} {\bibfnamefont {L.~M.}\ \bibnamefont
  {Robledo}},\ }\href {https://doi.org/10.1103/PhysRevC.78.014304} {\bibfield
  {journal} {\bibinfo  {journal} {Phys. Rev. C}\ }\textbf {\bibinfo {volume}
  {78}},\ \bibinfo {pages} {014304} (\bibinfo {year} {2008})}\BibitemShut
  {NoStop}%
\bibitem [{\citenamefont {Wang}\ \emph {et~al.}(2022)\citenamefont {Wang},
  \citenamefont {Sun},\ and\ \citenamefont {Zhou}}]{Wang2022_CPC46-024107}%
  \BibitemOpen
  \bibfield  {author} {\bibinfo {author} {\bibfnamefont {X.-Q.}\ \bibnamefont
  {Wang}}, \bibinfo {author} {\bibfnamefont {X.-X.}\ \bibnamefont {Sun}},\ and\
  \bibinfo {author} {\bibfnamefont {S.-G.}\ \bibnamefont {Zhou}},\ }\href
  {https://doi.org/10.1088/1674-1137/ac3904} {\bibfield  {journal} {\bibinfo
  {journal} {Chin. Phys. C}\ }\textbf {\bibinfo {volume} {46}},\ \bibinfo
  {pages} {024107} (\bibinfo {year} {2022})}\BibitemShut {NoStop}%
\bibitem [{\citenamefont {Rong}\ \emph {et~al.}(2023)\citenamefont {Rong},
  \citenamefont {Wu}, \citenamefont {Lu},\ and\ \citenamefont
  {Yao}}]{Rong2023_PLB840-137896}%
  \BibitemOpen
  \bibfield  {author} {\bibinfo {author} {\bibfnamefont {Y.-T.}\ \bibnamefont
  {Rong}}, \bibinfo {author} {\bibfnamefont {X.-Y.}\ \bibnamefont {Wu}},
  \bibinfo {author} {\bibfnamefont {B.-N.}\ \bibnamefont {Lu}},\ and\ \bibinfo
  {author} {\bibfnamefont {J.-M.}\ \bibnamefont {Yao}},\ }\href
  {https://doi.org/https://doi.org/10.1016/j.physletb.2023.137896} {\bibfield
  {journal} {\bibinfo  {journal} {Phys. Lett. B}\ }\textbf {\bibinfo {volume}
  {840}},\ \bibinfo {pages} {137896} (\bibinfo {year} {2023})}\BibitemShut
  {NoStop}%
\bibitem [{\citenamefont {Zheng}\ \emph {et~al.}(2014)\citenamefont {Zheng},
  \citenamefont {Xu}, \citenamefont {Shen}, \citenamefont {Liu}, \citenamefont
  {Wyss},\ and\ \citenamefont {Yan}}]{Zheng2014_PRC90-064309}%
  \BibitemOpen
  \bibfield  {author} {\bibinfo {author} {\bibfnamefont {S.~J.}\ \bibnamefont
  {Zheng}}, \bibinfo {author} {\bibfnamefont {F.~R.}\ \bibnamefont {Xu}},
  \bibinfo {author} {\bibfnamefont {S.~F.}\ \bibnamefont {Shen}}, \bibinfo
  {author} {\bibfnamefont {H.~L.}\ \bibnamefont {Liu}}, \bibinfo {author}
  {\bibfnamefont {R.}~\bibnamefont {Wyss}},\ and\ \bibinfo {author}
  {\bibfnamefont {Y.~P.}\ \bibnamefont {Yan}},\ }\href
  {https://doi.org/10.1103/PhysRevC.90.064309} {\bibfield  {journal} {\bibinfo
  {journal} {Phys. Rev. C}\ }\textbf {\bibinfo {volume} {90}},\ \bibinfo
  {pages} {064309} (\bibinfo {year} {2014})}\BibitemShut {NoStop}%
\bibitem [{\citenamefont {Lalazissis}\ \emph {et~al.}(1999)\citenamefont
  {Lalazissis}, \citenamefont {Raman},\ and\ \citenamefont
  {Ring}}]{Lalazissis1999_AtDataNuclDataTables71-1}%
  \BibitemOpen
  \bibfield  {author} {\bibinfo {author} {\bibfnamefont {G.}~\bibnamefont
  {Lalazissis}}, \bibinfo {author} {\bibfnamefont {S.}~\bibnamefont {Raman}},\
  and\ \bibinfo {author} {\bibfnamefont {P.}~\bibnamefont {Ring}},\ }\href
  {https://doi.org/https://doi.org/10.1006/adnd.1998.0795} {\bibfield
  {journal} {\bibinfo  {journal} {At. Data Nucl. Data Tables}\ }\textbf
  {\bibinfo {volume} {71}},\ \bibinfo {pages} {1} (\bibinfo {year}
  {1999})}\BibitemShut {NoStop}%
\bibitem [{\citenamefont {Zou}\ \emph {et~al.}(2010)\citenamefont {Zou},
  \citenamefont {Tian}, \citenamefont {Gu}, \citenamefont {Shen}, \citenamefont
  {Yao}, \citenamefont {Peng},\ and\ \citenamefont
  {Ma}}]{Zou2010_PRC82-024309}%
  \BibitemOpen
  \bibfield  {author} {\bibinfo {author} {\bibfnamefont {W.-H.}\ \bibnamefont
  {Zou}}, \bibinfo {author} {\bibfnamefont {Y.}~\bibnamefont {Tian}}, \bibinfo
  {author} {\bibfnamefont {J.-Z.}\ \bibnamefont {Gu}}, \bibinfo {author}
  {\bibfnamefont {S.-F.}\ \bibnamefont {Shen}}, \bibinfo {author}
  {\bibfnamefont {J.-M.}\ \bibnamefont {Yao}}, \bibinfo {author} {\bibfnamefont
  {B.-B.}\ \bibnamefont {Peng}},\ and\ \bibinfo {author} {\bibfnamefont
  {Z.-Y.}\ \bibnamefont {Ma}},\ }\href
  {https://doi.org/10.1103/PhysRevC.82.024309} {\bibfield  {journal} {\bibinfo
  {journal} {Phys. Rev. C}\ }\textbf {\bibinfo {volume} {82}},\ \bibinfo
  {pages} {024309} (\bibinfo {year} {2010})}\BibitemShut {NoStop}%
\bibitem [{\citenamefont {M$\ddot{\text{o}}$ller}\ \emph
  {et~al.}(2016)\citenamefont {M$\ddot{\text{o}}$ller}, \citenamefont {Sierk},
  \citenamefont {Ichikawa},\ and\ \citenamefont
  {Sagawa}}]{Moeller2016_ADNDT109-1}%
  \BibitemOpen
  \bibfield  {author} {\bibinfo {author} {\bibfnamefont {P.}~\bibnamefont
  {M$\ddot{\text{o}}$ller}}, \bibinfo {author} {\bibfnamefont {A.}~\bibnamefont
  {Sierk}}, \bibinfo {author} {\bibfnamefont {T.}~\bibnamefont {Ichikawa}},\
  and\ \bibinfo {author} {\bibfnamefont {H.}~\bibnamefont {Sagawa}},\ }\href
  {https://doi.org/https://doi.org/10.1016/j.adt.2015.10.002} {\bibfield
  {journal} {\bibinfo  {journal} {At. Data Nucl. Data Tables}\ }\textbf
  {\bibinfo {volume} {109-110}},\ \bibinfo {pages} {1} (\bibinfo {year}
  {2016})}\BibitemShut {NoStop}%
\bibitem [{\citenamefont {Rodr\'{i}guez}\ and\ \citenamefont
  {Egido}(2011)}]{Rodriguez2011_PLB705-255}%
  \BibitemOpen
  \bibfield  {author} {\bibinfo {author} {\bibfnamefont {T.~R.}\ \bibnamefont
  {Rodr\'{i}guez}}\ and\ \bibinfo {author} {\bibfnamefont {J.~L.}\ \bibnamefont
  {Egido}},\ }\href {https://doi.org/10.1016/j.physletb.2011.10.003} {\bibfield
   {journal} {\bibinfo  {journal} {Phys. Lett. B}\ }\textbf {\bibinfo {volume}
  {705}},\ \bibinfo {pages} {255} (\bibinfo {year} {2011})}\BibitemShut
  {NoStop}%
\bibitem [{\citenamefont {Zhang}\ \emph
  {et~al.}(2022{\natexlab{b}})\citenamefont {Zhang}, \citenamefont {Cheoun},
  \citenamefont {Choi}, \citenamefont {Chong}, \citenamefont {Dong},
  \citenamefont {Dong}, \citenamefont {Du}, \citenamefont {Geng}, \citenamefont
  {Ha}, \citenamefont {He}, \citenamefont {Heo}, \citenamefont {Ho},
  \citenamefont {In}, \citenamefont {Kim}, \citenamefont {Kim}, \citenamefont
  {Lee}, \citenamefont {Lee}, \citenamefont {Li}, \citenamefont {Li},
  \citenamefont {Luo}, \citenamefont {Meng}, \citenamefont {Mun}, \citenamefont
  {Niu}, \citenamefont {Pan}, \citenamefont {Papakonstantinou}, \citenamefont
  {Shang}, \citenamefont {Shen}, \citenamefont {Shen}, \citenamefont {Sun},
  \citenamefont {Sun}, \citenamefont {Tam}, \citenamefont {Thaivayongnou},
  \citenamefont {Wang}, \citenamefont {Wang}, \citenamefont {Wong},
  \citenamefont {Wu}, \citenamefont {Wu}, \citenamefont {Xia}, \citenamefont
  {Yan}, \citenamefont {Yeung}, \citenamefont {Yiu}, \citenamefont {Zhang},
  \citenamefont {Zhang}, \citenamefont {Zhang}, \citenamefont {Zhao},\ and\
  \citenamefont {Zhou}}]{Zhang2022_ADNDT144-101488}%
  \BibitemOpen
  \bibfield  {author} {\bibinfo {author} {\bibfnamefont {K.}~\bibnamefont
  {Zhang}}, \bibinfo {author} {\bibfnamefont {M.-K.}\ \bibnamefont {Cheoun}},
  \bibinfo {author} {\bibfnamefont {Y.-B.}\ \bibnamefont {Choi}}, \bibinfo
  {author} {\bibfnamefont {P.~S.}\ \bibnamefont {Chong}}, \bibinfo {author}
  {\bibfnamefont {J.}~\bibnamefont {Dong}}, \bibinfo {author} {\bibfnamefont
  {Z.}~\bibnamefont {Dong}}, \bibinfo {author} {\bibfnamefont {X.}~\bibnamefont
  {Du}}, \bibinfo {author} {\bibfnamefont {L.}~\bibnamefont {Geng}}, \bibinfo
  {author} {\bibfnamefont {E.}~\bibnamefont {Ha}}, \bibinfo {author}
  {\bibfnamefont {X.-T.}\ \bibnamefont {He}}, \bibinfo {author} {\bibfnamefont
  {C.}~\bibnamefont {Heo}}, \bibinfo {author} {\bibfnamefont {M.~C.}\
  \bibnamefont {Ho}}, \bibinfo {author} {\bibfnamefont {E.~J.}\ \bibnamefont
  {In}}, \bibinfo {author} {\bibfnamefont {S.}~\bibnamefont {Kim}}, \bibinfo
  {author} {\bibfnamefont {Y.}~\bibnamefont {Kim}}, \bibinfo {author}
  {\bibfnamefont {C.-H.}\ \bibnamefont {Lee}}, \bibinfo {author} {\bibfnamefont
  {J.}~\bibnamefont {Lee}}, \bibinfo {author} {\bibfnamefont {H.}~\bibnamefont
  {Li}}, \bibinfo {author} {\bibfnamefont {Z.}~\bibnamefont {Li}}, \bibinfo
  {author} {\bibfnamefont {T.}~\bibnamefont {Luo}}, \bibinfo {author}
  {\bibfnamefont {J.}~\bibnamefont {Meng}}, \bibinfo {author} {\bibfnamefont
  {M.-H.}\ \bibnamefont {Mun}}, \bibinfo {author} {\bibfnamefont
  {Z.}~\bibnamefont {Niu}}, \bibinfo {author} {\bibfnamefont {C.}~\bibnamefont
  {Pan}}, \bibinfo {author} {\bibfnamefont {P.}~\bibnamefont
  {Papakonstantinou}}, \bibinfo {author} {\bibfnamefont {X.}~\bibnamefont
  {Shang}}, \bibinfo {author} {\bibfnamefont {C.}~\bibnamefont {Shen}},
  \bibinfo {author} {\bibfnamefont {G.}~\bibnamefont {Shen}}, \bibinfo {author}
  {\bibfnamefont {W.}~\bibnamefont {Sun}}, \bibinfo {author} {\bibfnamefont
  {X.-X.}\ \bibnamefont {Sun}}, \bibinfo {author} {\bibfnamefont {C.~K.}\
  \bibnamefont {Tam}}, \bibinfo {author} {\bibnamefont {Thaivayongnou}},
  \bibinfo {author} {\bibfnamefont {C.}~\bibnamefont {Wang}}, \bibinfo {author}
  {\bibfnamefont {X.}~\bibnamefont {Wang}}, \bibinfo {author} {\bibfnamefont
  {S.~H.}\ \bibnamefont {Wong}}, \bibinfo {author} {\bibfnamefont
  {J.}~\bibnamefont {Wu}}, \bibinfo {author} {\bibfnamefont {X.}~\bibnamefont
  {Wu}}, \bibinfo {author} {\bibfnamefont {X.}~\bibnamefont {Xia}}, \bibinfo
  {author} {\bibfnamefont {Y.}~\bibnamefont {Yan}}, \bibinfo {author}
  {\bibfnamefont {R.~W.-Y.}\ \bibnamefont {Yeung}}, \bibinfo {author}
  {\bibfnamefont {T.~C.}\ \bibnamefont {Yiu}}, \bibinfo {author} {\bibfnamefont
  {S.}~\bibnamefont {Zhang}}, \bibinfo {author} {\bibfnamefont
  {W.}~\bibnamefont {Zhang}}, \bibinfo {author} {\bibfnamefont
  {X.}~\bibnamefont {Zhang}}, \bibinfo {author} {\bibfnamefont
  {Q.}~\bibnamefont {Zhao}},\ and\ \bibinfo {author} {\bibfnamefont {S.-G.}\
  \bibnamefont {Zhou}},\ }\href {https://doi.org/10.1016/j.adt.2022.101488}
  {\bibfield  {journal} {\bibinfo  {journal} {Atom. Data Nucl. Data Tables}\
  }\textbf {\bibinfo {volume} {144}},\ \bibinfo {pages} {101488} (\bibinfo
  {year} {2022}{\natexlab{b}})}\BibitemShut {NoStop}%
\bibitem [{\citenamefont {Zberecki}\ \emph {et~al.}(2006)\citenamefont
  {Zberecki}, \citenamefont {Magierski}, \citenamefont {Heenen},\ and\
  \citenamefont {Schunck}}]{Zberecki2006_PRC74-051302R}%
  \BibitemOpen
  \bibfield  {author} {\bibinfo {author} {\bibfnamefont {K.}~\bibnamefont
  {Zberecki}}, \bibinfo {author} {\bibfnamefont {P.}~\bibnamefont {Magierski}},
  \bibinfo {author} {\bibfnamefont {P.-H.}\ \bibnamefont {Heenen}},\ and\
  \bibinfo {author} {\bibfnamefont {N.}~\bibnamefont {Schunck}},\ }\href
  {https://doi.org/10.1103/PhysRevC.74.051302} {\bibfield  {journal} {\bibinfo
  {journal} {Phys. Rev. C}\ }\textbf {\bibinfo {volume} {74}},\ \bibinfo
  {pages} {051302(R)} (\bibinfo {year} {2006})}\BibitemShut {NoStop}%
\bibitem [{\citenamefont {Lalazissis}\ \emph {et~al.}(1997)\citenamefont
  {Lalazissis}, \citenamefont {K\"onig},\ and\ \citenamefont
  {Ring}}]{Lalazissis1997_PRC55-540}%
  \BibitemOpen
  \bibfield  {author} {\bibinfo {author} {\bibfnamefont {G.~A.}\ \bibnamefont
  {Lalazissis}}, \bibinfo {author} {\bibfnamefont {J.}~\bibnamefont
  {K\"onig}},\ and\ \bibinfo {author} {\bibfnamefont {P.}~\bibnamefont
  {Ring}},\ }\href {https://doi.org/10.1103/PhysRevC.55.540} {\bibfield
  {journal} {\bibinfo  {journal} {Phys. Rev. C}\ }\textbf {\bibinfo {volume}
  {55}},\ \bibinfo {pages} {540} (\bibinfo {year} {1997})}\BibitemShut
  {NoStop}%
\bibitem [{\citenamefont {Zhao}\ \emph {et~al.}(2010)\citenamefont {Zhao},
  \citenamefont {Li}, \citenamefont {Yao},\ and\ \citenamefont
  {Meng}}]{Zhao2010_PRC82-054319}%
  \BibitemOpen
  \bibfield  {author} {\bibinfo {author} {\bibfnamefont {P.~W.}\ \bibnamefont
  {Zhao}}, \bibinfo {author} {\bibfnamefont {Z.~P.}\ \bibnamefont {Li}},
  \bibinfo {author} {\bibfnamefont {J.~M.}\ \bibnamefont {Yao}},\ and\ \bibinfo
  {author} {\bibfnamefont {J.}~\bibnamefont {Meng}},\ }\href
  {https://doi.org/10.1103/PhysRevC.82.054319} {\bibfield  {journal} {\bibinfo
  {journal} {Phys. Rev. C}\ }\textbf {\bibinfo {volume} {82}},\ \bibinfo
  {pages} {054319} (\bibinfo {year} {2010})}\BibitemShut {NoStop}%
\bibitem [{\citenamefont {Nik\ifmmode \check{s}\else
  \v{s}\fi{}i\ifmmode~\acute{c}\else \'{c}\fi{}}\ \emph
  {et~al.}(2008)\citenamefont {Nik\ifmmode \check{s}\else
  \v{s}\fi{}i\ifmmode~\acute{c}\else \'{c}\fi{}}, \citenamefont {Vretenar},\
  and\ \citenamefont {Ring}}]{Niksic2008_PRC78-034318}%
  \BibitemOpen
  \bibfield  {author} {\bibinfo {author} {\bibfnamefont {T.}~\bibnamefont
  {Nik\ifmmode \check{s}\else \v{s}\fi{}i\ifmmode~\acute{c}\else \'{c}\fi{}}},
  \bibinfo {author} {\bibfnamefont {D.}~\bibnamefont {Vretenar}},\ and\
  \bibinfo {author} {\bibfnamefont {P.}~\bibnamefont {Ring}},\ }\href
  {https://doi.org/10.1103/PhysRevC.78.034318} {\bibfield  {journal} {\bibinfo
  {journal} {Phys. Rev. C}\ }\textbf {\bibinfo {volume} {78}},\ \bibinfo
  {pages} {034318} (\bibinfo {year} {2008})}\BibitemShut {NoStop}%
\bibitem [{\citenamefont {Ajimura}\ \emph {et~al.}(2001)\citenamefont
  {Ajimura}, \citenamefont {Hayakawa}, \citenamefont {Kishimoto}, \citenamefont
  {Kohri}, \citenamefont {Matsuoka}, \citenamefont {Minami}, \citenamefont
  {Mori}, \citenamefont {Morikubo}, \citenamefont {Saji}, \citenamefont
  {Sakaguchi}, \citenamefont {Shimizu}, \citenamefont {Sumihama}, \citenamefont
  {Chrien}, \citenamefont {May}, \citenamefont {Pile}, \citenamefont {Rusek},
  \citenamefont {Sutter}, \citenamefont {Eugenio}, \citenamefont {Franklin},
  \citenamefont {Khaustov}, \citenamefont {Paschke}, \citenamefont {Quinn},
  \citenamefont {Schumacher}, \citenamefont {Franz}, \citenamefont {Fukuda},
  \citenamefont {Noumi}, \citenamefont {Outa}, \citenamefont {Gan},
  \citenamefont {Tang}, \citenamefont {Yuan}, \citenamefont {Tamura},
  \citenamefont {Nakano}, \citenamefont {Tamagawa}, \citenamefont {Tanida},\
  and\ \citenamefont {Sawafta}}]{Ajimura2001_PRL86-4255}%
  \BibitemOpen
  \bibfield  {author} {\bibinfo {author} {\bibfnamefont {S.}~\bibnamefont
  {Ajimura}}, \bibinfo {author} {\bibfnamefont {H.}~\bibnamefont {Hayakawa}},
  \bibinfo {author} {\bibfnamefont {T.}~\bibnamefont {Kishimoto}}, \bibinfo
  {author} {\bibfnamefont {H.}~\bibnamefont {Kohri}}, \bibinfo {author}
  {\bibfnamefont {K.}~\bibnamefont {Matsuoka}}, \bibinfo {author}
  {\bibfnamefont {S.}~\bibnamefont {Minami}}, \bibinfo {author} {\bibfnamefont
  {T.}~\bibnamefont {Mori}}, \bibinfo {author} {\bibfnamefont {K.}~\bibnamefont
  {Morikubo}}, \bibinfo {author} {\bibfnamefont {E.}~\bibnamefont {Saji}},
  \bibinfo {author} {\bibfnamefont {A.}~\bibnamefont {Sakaguchi}}, \bibinfo
  {author} {\bibfnamefont {Y.}~\bibnamefont {Shimizu}}, \bibinfo {author}
  {\bibfnamefont {M.}~\bibnamefont {Sumihama}}, \bibinfo {author}
  {\bibfnamefont {R.~E.}\ \bibnamefont {Chrien}}, \bibinfo {author}
  {\bibfnamefont {M.}~\bibnamefont {May}}, \bibinfo {author} {\bibfnamefont
  {P.}~\bibnamefont {Pile}}, \bibinfo {author} {\bibfnamefont {A.}~\bibnamefont
  {Rusek}}, \bibinfo {author} {\bibfnamefont {R.}~\bibnamefont {Sutter}},
  \bibinfo {author} {\bibfnamefont {P.}~\bibnamefont {Eugenio}}, \bibinfo
  {author} {\bibfnamefont {G.}~\bibnamefont {Franklin}}, \bibinfo {author}
  {\bibfnamefont {P.}~\bibnamefont {Khaustov}}, \bibinfo {author}
  {\bibfnamefont {K.}~\bibnamefont {Paschke}}, \bibinfo {author} {\bibfnamefont
  {B.~P.}\ \bibnamefont {Quinn}}, \bibinfo {author} {\bibfnamefont {R.~A.}\
  \bibnamefont {Schumacher}}, \bibinfo {author} {\bibfnamefont
  {J.}~\bibnamefont {Franz}}, \bibinfo {author} {\bibfnamefont
  {T.}~\bibnamefont {Fukuda}}, \bibinfo {author} {\bibfnamefont
  {H.}~\bibnamefont {Noumi}}, \bibinfo {author} {\bibfnamefont
  {H.}~\bibnamefont {Outa}}, \bibinfo {author} {\bibfnamefont {L.}~\bibnamefont
  {Gan}}, \bibinfo {author} {\bibfnamefont {L.}~\bibnamefont {Tang}}, \bibinfo
  {author} {\bibfnamefont {L.}~\bibnamefont {Yuan}}, \bibinfo {author}
  {\bibfnamefont {H.}~\bibnamefont {Tamura}}, \bibinfo {author} {\bibfnamefont
  {J.}~\bibnamefont {Nakano}}, \bibinfo {author} {\bibfnamefont
  {T.}~\bibnamefont {Tamagawa}}, \bibinfo {author} {\bibfnamefont
  {K.}~\bibnamefont {Tanida}},\ and\ \bibinfo {author} {\bibfnamefont
  {R.}~\bibnamefont {Sawafta}},\ }\href
  {https://doi.org/10.1103/PhysRevLett.86.4255} {\bibfield  {journal} {\bibinfo
   {journal} {Phys. Rev. Lett.}\ }\textbf {\bibinfo {volume} {86}},\ \bibinfo
  {pages} {4255} (\bibinfo {year} {2001})}\BibitemShut {NoStop}%
\bibitem [{\citenamefont {Zhou}\ \emph {et~al.}(2016)\citenamefont {Zhou},
  \citenamefont {Hiyama},\ and\ \citenamefont
  {Sagawa}}]{Zhou2016_PRC94-024331}%
  \BibitemOpen
  \bibfield  {author} {\bibinfo {author} {\bibfnamefont {X.-R.}\ \bibnamefont
  {Zhou}}, \bibinfo {author} {\bibfnamefont {E.}~\bibnamefont {Hiyama}},\ and\
  \bibinfo {author} {\bibfnamefont {H.}~\bibnamefont {Sagawa}},\ }\href
  {https://doi.org/10.1103/PhysRevC.94.024331} {\bibfield  {journal} {\bibinfo
  {journal} {Phys. Rev. C}\ }\textbf {\bibinfo {volume} {94}},\ \bibinfo
  {pages} {024331} (\bibinfo {year} {2016})}\BibitemShut {NoStop}%
\bibitem [{\citenamefont {Wu}\ \emph {et~al.}(2017)\citenamefont {Wu},
  \citenamefont {Mei}, \citenamefont {Yao},\ and\ \citenamefont
  {Zhou}}]{Wu2017_PRC95-034309}%
  \BibitemOpen
  \bibfield  {author} {\bibinfo {author} {\bibfnamefont {X.~Y.}\ \bibnamefont
  {Wu}}, \bibinfo {author} {\bibfnamefont {H.}~\bibnamefont {Mei}}, \bibinfo
  {author} {\bibfnamefont {J.~M.}\ \bibnamefont {Yao}},\ and\ \bibinfo {author}
  {\bibfnamefont {X.-R.}\ \bibnamefont {Zhou}},\ }\href
  {https://doi.org/10.1103/PhysRevC.95.034309} {\bibfield  {journal} {\bibinfo
  {journal} {Phys. Rev. C}\ }\textbf {\bibinfo {volume} {95}},\ \bibinfo
  {pages} {034309} (\bibinfo {year} {2017})}\BibitemShut {NoStop}%
\end{thebibliography}%

\end{document}